\documentclass[aps,prl,floatfix,twocolumn,superscriptaddress,showpacs,epsf]{revtex4-1}

\usepackage[USenglish]{babel}
\usepackage{amsmath,amssymb,amsfonts}
\usepackage{epstopdf}
\usepackage{epsfig} 
\usepackage{graphicx}
\usepackage{subfigure}
\usepackage{import}
\usepackage{color}
\usepackage{url}

\usepackage{changes}

\usepackage{mathtools}

\allowdisplaybreaks

\usepackage[colorlinks=true,linkcolor=blue,citecolor=red]{hyperref}

\makeatletter
\newcommand\resetchangescolor[1]{%
  \setkeys{Changes@definechangesauthor}{color=#1}%
  \expandafter%
  \let\csname Changes@AuthorColor\endcsname=\Changes@definechangesauthor@color%
  \colorlet{Changes@Color}{\@nameuse{Changes@AuthorColor}}%
}
\makeatother

\makeatletter
\renewcommand\tableofcontents{\@starttoc{toc}}
\makeatother
\def\bcen{\begin{center}}
\def\ecen{\end{center}}

\def\a{\alpha}       \def\b{\beta}   \def\g{\gamma}   \def\d{\delta} 
\def\e{\varepsilon}  \def\z{\zeta}   \def\h{\eta}     \def\th{\theta}
\def\k{\kappa}       \def\l{\lambda} \def\m{\mu}      \def\n{\nu}
\def\x{\xi}          \def\p{\pi}     \def\r{\rho}     \def\s{\sigma}
\def\t{\tau}         \def\f{\varphi} \def\ph{\varphi} \def\c{\chi}
\def\ps{\pi}        \def\y{\upsilon}\def\o{\omega}   \def\si{\varsigma}
\def\G{\Gamma}       \def\D{\Delta}  \def\Th{\Theta}  \def\L{\Lambda}  
\def\X{\Xi}          \def\P{\Pi}     \def\Si{\Sigma}  \def\F{\Phi}    
\def\Ps{\Psi}        \def\O{\Omega}  \def\Y{\Upsilon} \def\lg{\langle}

\def\PP{{\cal P}}\def\EE{{\cal E}}\def\MM{{\cal M}} \def\VV{{\cal V}}
\def\CC{{\cal C}}\def\FF{{\cal F}}\def\HH{{\cal H}}\def\WW{{\cal W}}
\def\TT{{\cal T}}\def\NN{{\cal N}}\def\BB{{\cal B}} \def\II{{\cal I}}
\def\RR{{\cal R}}\def\LL{{\cal L}}\def\JJ{{\cal J}} \def\OO{{\cal O}}
\def\DD{{\cal D}}
\def\AAA{{\cal A}}
\def\GG{{\cal G}} \def\SS{{\cal S}}
\def\ZZ{{\cal Z}} \def\UU{{\cal U}}
\def\SB{{\cal S}{\cal B}}
\def\aa{{\V \a}}
\def\hh{{\V h}}\def\HHH{{\V H}}
\def\nn{{\V \n}}\def\pp{{\V p}}\def\mm{{\V m}}\def\qq{{\bf q}}
\def\RRR{\mathbb{R}} \def\CCC{\mathbb{C}} \def\NNN{\mathbb{N}}
\def\ZZZ{\mathbb{Z}} 
\def\QQQ{\hbox{\msytw Q}}

\def\AA{\buildrel_{\circ}\over{\mathrm{A}}}

\def\rg{\rangle}
 \def\ul{\underline}
\def\eg{\mbox{\it e.g.\ }}  \def\ie{\mbox{\it i.e.\ }}
\def\=={\equiv} \def\defi{{\buildrel def \over =}}
\def\lft{\left} \def\rgt{\right} \def\dpr{\partial} \def\der{{\rm d}}
\def\us{\underline \s} \def\ue{{\underline \e}} \def\la{\left\langle}
\def\ra{\right\rangle} 
\def\qed{\raise1pt\hbox{\vrule height5pt width5pt depth0pt}}
\def\iome{i\omega_n} \def\iom{i\omega} \def\iom#1{i\omega_{#1}}
\def\iomn{i\omega_n}
\def\epsk{\epsilon({\bf k})} \def\Ga{\Gamma_{\alpha}}
\def\Seff{S_{eff}}  \def\dinf{$d\rightarrow\infty\,$}
\def\cG0{{\cal G}_0} 
\def\cG{{\cal G}}  \def\cU{{\cal U}}  \def\cS{{\cal S}}
\def\divnum{\frac{1}{N_s}} \def\vac{|\mbox{vac}\rangle}
\def\intR{\int_{-\infty}^{+\infty}} \def\intbeta{\int_{0}^{\beta}}
\def\spinup{\uparrow} \def\spindown{\downarrow} 
\def\up{\uparrow} \def\down{\downarrow} \def\dw{\downarrow}
\def\vk{{\bf k}} \def\qa{{\bf q}} \def\vQ{{\bf Q}}
\def\bk{{\bf k}}\def\bR{{\bf R}}
\def\bq{{\bf q}}
\def\ka{{\bf k} \alpha} 
\def\vr{{\bf r}} \def\q{{\bf q}}  \def\R{{\bf R}}  \def\vR{{\bf R}}
\def\Ak{{\bf A}} \def\Akt{{\bf A}(t)} \def\Ek{{\mathbf E}}
\def\kp{\bbox{k'}} \def\hc{\mbox{h.c.}} \def\Im{\mbox{Im}}
\def\ie{\hbox{\it i.e.\ }} \def\eg{\mbox{\it e.g.\ }}
\def\ie{\mbox{\it i.e.\ }} \def\=={\equiv}
\def\defi{{\buildrel def \over =}} \def\nt{\widetilde{n}}
\def\Im{{\rm Im}} \def\Re{{\rm Re}} \def\Tr{{\rm Tr}\,}
\def\det{{\rm det}\,} \def\ep0{\epsilon_{p}} \def\ed0{\epsilon_{f}}
\def\tpd{V_{fp}} \def\unmezzo{\frac{1}{2}}
\def\ispin{\{\underline{s}\}}
\def\ispinp{\{\underline{s'}\}}

\def\dt{\Delta \tau}

\def\be{\begin{equation}}
\def\ee{\end{equation}}

\def\cc{c^{\dagger}}
\def\ca{c^{\phantom{\dagger}}}
\def\dc{d^{\dagger}}
\def\da{d^{\phantom{\dagger}}}
\def\ac{a^{\dagger}}
\def\aa{a^{\phantom{\dagger}}}

\def\tc{\tilde{c}^{\dagger}}
\def\ta{\tilde{c}^{\phantom{\dagger}}}

\def\abcd{\alpha \beta \gamma \delta}

\newcommand{\quave}[1]{\langle{#1}\rangle}

\newcommand{\ms}[1]{{\color{black}{#1}}}

\def\pf{\psi}
\def\pfs{\psi^*}

\def\pb{\phi}
\def\pbs{\phi^*}

\def\Hsys{H_{\text{sys}}}
\def\Hbath{H_{\text{bath}}}
\def\Hsb{H_{\text{s-b}}}
\def\rsb{\rho_{I}}
\def\drsb{\dot{\rho}_{I}(t)}

\usepackage{fancyhdr}
\begin{document}
\resetchangescolor{red}

\author{Giacomo Mazza}
\affiliation{Dipartimento di Fisica dell'Universit\`a di Pisa, Largo Bruno Pontecorvo 3, I-56127 Pisa, Italy}
\affiliation{Department of Quantum Matter Physics, University of Geneva, Quai Ernest-Ansermet 24, 1211 Geneva, Switzerland}
\author{Marco Schir\`o}
\affiliation{JEIP, UAR 3573 CNRS, Coll\`{e}ge de France, PSL Research University, 11 Place Marcelin Berthelot, 75321 Paris Cedex 05, France}

\title{Dissipative Dynamics of a Fermionic Superfluid with Two-Body Losses}

\begin{abstract}
We study the dissipative dynamics of a fermionic superfluid in presence of two-body losses. We use a variational approach for the Lindblad dynamics and obtain dynamical equations for Anderson's pseudo-spins where dissipation enters as a complex pairing interaction as well as effective, density-dependent, single particle losses which break the conservation of the pseudo-spin norm. We show that this latter has key consequences on the dynamical behavior of the system. In the case of a sudden switching of two-body losses we show that the superfluid order parameter decays much faster than then particle density at short times and eventually slows-down, setting into a power-law decay at longer time scales driven by the depletion of the system.  We then consider a  quench of pairing interaction, leading to coherent oscillations in the unitary case, followed by the switching of the dissipation. We show that losses 
affect the dynamical BCS synchronization 
by introducing not only damping but also a renormalization of the frequency of coherent oscillations, 
which depends non-linearly from the rate of the two-body losses.
\end{abstract}

\maketitle

\paragraph{Introduction -} 

\ms{The nonequilibrium dynamics of superfluids and superconductors has attracted fresh new interest in recent years. 
Non-linear optical spectroscopy and manipulation of collective modes in the superconducting phase have been demonstrated~\cite{matsunaga2014light,shimano2020higgs} together with reports of light-induced superconductivity in variety of materials~\cite{Fausti_Science11,MitranoNature16,Buzzi2020photomolecular,budden2021evidence}, among the most striking demonstration of light control of quantum matter and Floquet-engineering~\cite{giannetti2016ultrafast,Oka2018,delatorre2021nonthermal}. In atomic physics the realization of fermionic superfluids ~\cite{regal2004observation,zwierlein2004condensation,bartenstein2004crossover} has led to the investigation of different 
dynamical phenomena, such  the spectroscopy of driven superfluids~\cite{behrle2019higgs}.
These experimental developments have stimulated theoretical interest on the subject of dynamics in superfluids and superconductors~
\cite{BarankovLevitovSpivakPRL04,BarankovLevitovPRL06,YuzbashyanEtAlPRB05,YuzbashyanEtAlPRL06,GurariePRL09,foster2014quench,
kurkjian2019pair,mazza2017nonequilibrium,babadi2017theory,nava2018cooling,li2020eta,peronaci2020enhancement,peronaci2015transient}. }
 In most cases, theoretical investigations of these phenomena have focused on the dynamics of closed isolated systems. Dissipation is however  not only unavoidable in realistic experimental contexts, such as in the solid-state, but can sometime be controlled with high-degree of flexibility, as in certain ultracold atoms experiments, and used as a tool to control the dynamical long-time behavior of the system. Dissipative quantum many-body systems represent a fresh platform where novel dynamical phenomena and phase transition can appear as result of the competition between unitary evolution and dissipative couplings~\cite{poletti2012interaction,poletti2013emergence,sciolla2015twotime,pan2020nonhermitian}.

A particularly interesting scenario is realized when dissipation has a genuine many-body character, since it involves correlated processes such as heating due to stimulated emission~\cite{gerbier2010heating,bouganne2020,tindall2019heating}, spontaneous emission~\cite{nakagawa2021etapairing} or two-particle losses~\cite{kantian2009atomic,Syassen1329,TomitaEtAlScienceAdv17,Sponselee_2018,honda2022observation,wang2022complex,zhou2021effective}. 
These types of dissipative inelastic scattering processes naturally arise for example in experiments with ultracold fermions made of Alkali-Earth atoms~\cite{sandner2011spatial,zhang2015orbital,pagano2015strongly}. Their role for the dynamics has recently attracted large interest in the context of Dicke states~\cite{fossfeig2012steady,Sponselee_2018,rosso2021onedimensional} and Quantum Zeno Effect (QZE)~\cite{misra1977a} where the effective dissipation decreases as the loss rate is increased~\cite{garcia-ripoll2009,zhu2014suppressing,Froml2019,NakagawaEtAlPRL20,rossini2021strong,scarlatella2021dynamical,Biella2021manybodyquantumzeno,
rosso2022the,rosso2022adynamical,secli2022steady}.  
\ms{The effect of two-body losses on the dynamics of superfluids and superconductors is particularly intriguing, since  dissipation here affects directly the degrees of freedom involved in the condensate and could for example couple non trivially to its collective modes or induce non trivial responses which are not expected for single particle dissipative processes.}

In this Letter we study the dissipative dynamics of a fermionic superfluid, 
modelled as an attractive Hubbard model~\cite{mitra2018quantum} 
in presence of weak local two-body losses. 
Recent works in this context have focused on 
simplified descriptions of dissipation in terms of 
 a non-Hermitian Bardeen-Cooper-Schrieffer (BCS) problem~\cite{YamamotoEtAlPRL19,iskin2021nonhermitian} or an effective unitary dynamics with complex pairing potential~\cite{yamamoto2021collective}.  
Here, using a variational approach for Lindblad dynamics, we 
show that a complete dissipative BCS theory includes an effective, 
density-dependent, single particle loss term, which corresponds to 
decoupling the two-body losses in the particle-particle and 
particle-hole channels. We show that this term completely controls the long-time dynamics 
of the system \ms{leading to a universal power-law decay of particle density and 
to a crossover in the superfluid order parameter from a short-time exponential decay to a long-time power law decay controlled by the depletion of the system.}
\ms{Furthermore we show that a weak dissipation has a dramatic effect on the BCS synchronization dynamics~\cite{BarankovLevitovPRL06} whose frequency of coherent oscillations is strongly renormalised. We understand this effect as arising from a weak-breaking of Anderson pseudospins length and construct a dissipative soliton solution which qualitatively captures the observed frequency renormalization.}
Our results can be experimentally tested in experiments with ultracold fermionic superfluids~\cite{kwon2020strongly,delpace2021tunneling}, where two-body losses can be introduced through photoassociation~\cite{TomitaEtAlScienceAdv17,honda2022observation} as well as cavity QED simulators of nonequilibrium superfluidity~\cite{muniz2020exploring,lewis2021cavity}.

\paragraph{Model - } We consider a system of spinful fermions hopping on a lattice, in presence of a local pairing interaction as described by the attractive Hubbard model whose  Hamiltonian reads
\begin{align}
H=\sum_{<ij>\s} t_{ij} \cc_{i \s} \ca_{j\s} - |U| \sum_i n_{i \up} n_{i \downarrow}
\end{align}
where $-|U|$ is the attraction and the $t_{ij}$ the nearest neighbor hopping.
\ms{The hopping gives rise to a single particle band of width $W$.
For simplicity, we consider a band characterized by a
flat density of states.
Different choices do not affect in any qualitative way our results as long 
as the density of states is non-singular.}
This model has been studied in thermal equilibrium~\cite{Toschi_2005} in the context of the BCS to BEC superfluidity crossover~\cite{bourdel2004experimental,CHEN20051,biss2022excitation,mazza_interface_BCSBEC_2021}, while its unitary dynamics has received attention recently and revealed a variety of dynamical phase transitions~\cite{sentef2017theory,mazza2017fromsudden,seibold2020nonequilibrium,
ojeda2019fate,mazza2012dynamical,seibold2022adiabatic,collado2205} . 
Here we focus on an open quantum system setting in which the evolution of the system density matrix $\rho(t)$ is described by a Lindblad master equation~\cite{breuerPetruccione2007}, $(\hbar=1)$,
\begin{equation}\label{eqn:lindblad}
\partial_t \rho = - i[H,\rho]+\sum_{i} \left(L_{i} \rho L_{i}^{\dagger}
-\frac{1}{2}\left\{L^{\dagger}_{i} L_{i},\rho \right\}\right)\,.
\end{equation}
with local, on-site,  jump operators describing Markovian dissipation. 
Here we consider dissipative processes in which pairs of fermions on the same site and with opposite spins escape from the system to the environment, leading to a jump operator of the form  {$L_i =\sqrt{\G} \ca_{i \down} \ca_{i \up}$}. The resulting dissipative dynamics does not conserve the total number of particles. In absence of any driving term to counterbalance the loss of particles into the environment the system evolves at long times towards the zero density limit. We note that two-body losses conserve instead the total spin which would prevent from reaching complete depletion~\cite{rosso2021onedimensional}, unless the system is initially prepared in a total singlet state as it is our case here. While the stationary state properties of the model are therefore trivial its depletion dynamics can still reveal intriguing features and give rise to different dynamical regimes, as we are going to discuss.

\paragraph{Method  - } To study the dynamics of the system we use a time-dependent variational approach. While for unitary system the Dirac's variational principle is a standard and much used result, both for gaussian and for correlated wave functions,  its generalisation to the open system case pose some challenges. Recent work~\cite{weimer2015variational} has proposed a variational principle for the stationary state which is however not of direct use here, where the long-time limit is the vacuum. To focus on dynamics we proceed along a different line, directly inspired by work on unitary quantum dynamics. We note that stating that a density matrix $\rho$ evolves according to Eq.~\ref{eqn:lindblad} is equivalent to say that the functional $S[\rho_0,\rho_{\mathrm{aux}}]= \int\, dt \Tr \left[ \rho_{\mathrm{aux}} (i \partial_{t} \rho_0 -  {\cal L}[\rho_0] ) \right] $ is stationary with respect to any given density matrix 
$\rho_\mathrm{aux}$. 
Using this condition on a Gaussian density matrix $\rho_0$ for which Wick's theorem applies, including normal and anomalous contractions, allows us to obtain the following variational dynamics~\cite{SM}
\begin{equation}
\partial_t \r_0 = -i\left[ \tilde{H}_{BCS},\rho_0 \right] + \G  \frac{n}{2} \sum_\s {\cal L}^{\mathrm{1p-loss}}_\s[\rho_0] 
\label{eq:effective_lindblad}
\end{equation}
{which takes the form of an effective Lindblad master equation. Here the unitary part comes from the usual BCS mean-field Hamiltonian  plus an imaginary pairing field 
$ i\Gamma$ 
\begin{equation}
\tilde{H}_{BCS} = H_{BCS} + i \G \sum_i\left(\D \ca_{i \down} \ca_{i \up} -  \D^* \cc_{i \up} \cc_{i \down}\right)\,
\end{equation}
where $\Delta$ is the superfluid order parameter
$$
\Delta(t)=\frac{1}{V}\sum_{\bk}\mbox{Tr}\left(\rho_0 c^{\dagger}_{\bk \uparrow} c^{\dagger}_{-\bk \downarrow}\right)
$$
while the dissipative part ${\cal L}^{\mathrm{1p-loss}}_\s[\rho_0] $ in Eq.~(\ref{eq:effective_lindblad}) contains effective single-particle losses of strength $\Gamma_{\rm eff}=n(t)\Gamma$, with  $n(t)=\frac{1}{V}\sum_{i\s}\mbox{Tr}\rho_0 n_{i\s} $ the time-dependent particle density.}
The variational dynamics associated to the above effective Lindbladian reads
\begin{align}\label{eq:sigmax}
\dot{\sigma}_{\bk}^x &=-2\varepsilon_{\bk}\sigma^y_{\bk}+2\mbox{Im}(\Phi)\sigma^z_{\bk}-\Gamma n\sigma^x_{\bk}\\
\label{eq:sigmay}
\dot{\sigma}_{\bk}^y &=2\varepsilon_{\bk}\sigma^x_{\bk}-2\mbox{Re}(\Phi)\sigma^z_{\bk}-\Gamma n\sigma^y_{\bk}\\
\dot{\sigma}_{\bk}^z &=2\mbox{Re}(\Phi)\sigma^y_{\bk}-2\mbox{Im}(\Phi)\sigma^x_{\bk}-\Gamma n\left(\sigma^z_{\bk}+1\right)\label{eq:sigmaz}
\end{align}
where we have introduced the Anderson's pseudo-spin $\sigma^{\alpha}_{\bk}=
\mbox{Tr}\left(\rho_0 \Psi^{\dagger}_{\bk}\sigma^{\alpha} \Psi_{-\bk}\right)$ with $\sigma^{\alpha=x,y,z}$ given by the Pauli matrices, where $\varepsilon_{\bk}$ is the bare energy dispersion of the lattice, and {$\Phi(t)=\left(-\vert U\vert+i\Gamma\right)\Delta(t)$ is the self-consistent pairing field}. This dynamics describes the competition between precession of Anderson's pseudo-spin around an effective magnetic field, as in the unitary case, and losses-induced decoherence towards the steady state $\sigma^x_{\bk}=\sigma^y_{\bk}=0$ and $\sigma^z_{\bk}=-1$, corresponding to vanishing order parameter and density. We note that the length of the pseudo spin $\mathcal{S}=\sum_{\alpha\bk}\left(\sigma^{\alpha}_{\bk}\right)^2$ is \emph{not} conserved due to the presence of the single particle loss term proportional to the density. 
Furthermore the purity of the variational state $P=\mbox{Tr}\left(\rho_0^2\right)$ is also not conserved, as expected for a dissipative Lindblad dynamics. The dynamical equations above differ therefore from those that can be obtained by Hubbard-Stratonovich decoupling~\cite{yamamoto2021collective}, which essentially take the form of a unitary dynamics with a complex pairing term $U+i\Gamma$. This difference arises  due to the presence of the effective single particle loss term in Eq.~(\ref{eq:effective_lindblad}), that couples the Keldysh contours. We will discuss below the consequences of this term for the physics of the problem. We note that instead the equations above coincide with those that can be obtained through a direct mean-field decoupling of Hamiltonian and dissipator, including both contributions coming from particle-particle and particle-hole channels.
\begin{figure}[t!]
\includegraphics[width=\columnwidth]{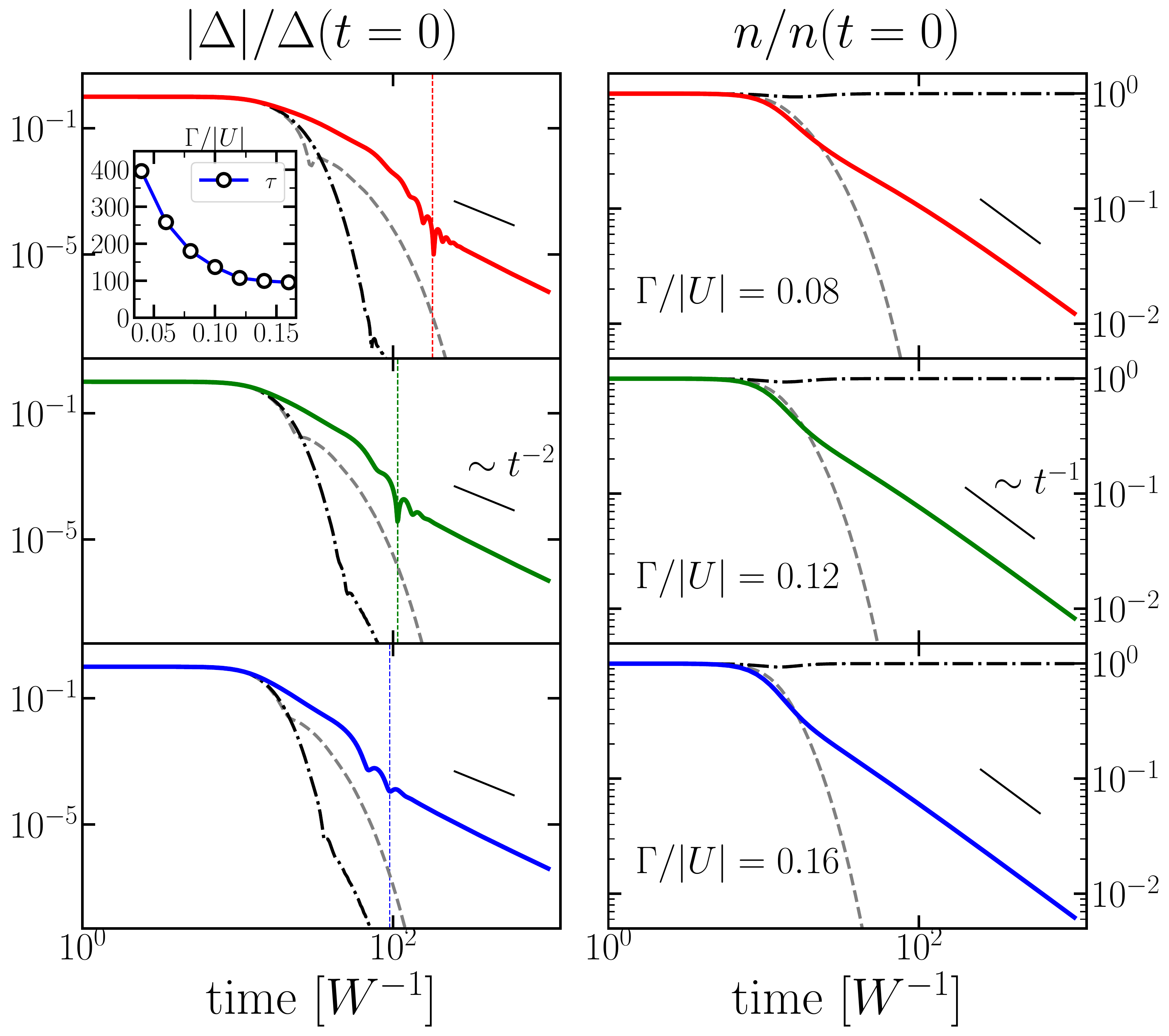}
	\caption{(Left panels) Dynamics of the order parameter  
        after 
	a sudden quench of two-body losses, for three values $\G/|U| = 0.08$, 
	$\G/|U| = 0.12$ $\G/|U| = 0.16$  from top to bottom and $|U|/W = 1.0.$.
	(Right panels) Dynamics of the particle density for the same parameters used in the 
	left panels. 
	The full lines mark the $\sim t^{-1}$ behavior for the density, 
	and  $\sim t^{-2}$ for the order parameter.
	Dot-dashed lines represent same quantities in the presence of an 
	additional single-particle pump that keeps the density constant.
	Dashed lines represent the dissipative dynamics considering single-particle 
	losses in place of the two-body ones. 
	The inset shows the time scale of the crossover from the 
	exponential to power law behavior of the order parameter amplitude 
	(vertical lines in left panels.)
	}
	\label{fig1}
\end{figure}

\paragraph{Results - Dissipation quench} We begin our discussion from the dynamics after a sudden switching of the two-body losses $\Gamma$, starting from the ground-state of the attractive Hubbard model with $|U|/W=1.0.$

 In Fig.~\ref{fig1} we plot the time evolution of the order parameter $\Delta(t)$ and particle density $n(t)$ for different values of the dissipation measured with respect to the interaction $\Gamma/|U|$. We see in the right panels that the density remains constant at short times while above a time scale 
 which depends weakly on the loss rate it displays a power-law decay towards zero, 
 corresponding to the vacuum state,  with an exponent $\sim t^{-1}$ which is independent of  $\Gamma$.  
 On the other hand the dynamics of the superfluid order parameter $\Delta(t)$ is richer 
and shows a crossover from an exponential decay at short times
 followed by a slower power law decay on longer time scales. 
\ms{Importantly, we see that the decay of the order parameter is faster than the density and compatible with a power law decay
$\vert\Delta\vert\sim 1/t^2$, whose origin we will discuss below.
 We argue that the crossover in the dynamics of the order parameter, occurring on 
 a time scale $\tau$ which decreases with $\Gamma$ (see inset in Fig.~\ref{fig1}), 
  is a key dynamical signature of a dissipative superfluid with two-body 
  losses, and it is controlled by the slow depletion of the system.
For comparison, we show in Fig.~\ref{fig1} the behaviour obtained when
only single-particle losses (dashed lines) are present, or in the presence of an additional
single-particle pump to keep the density constant in time (dot-dashed lines)~\cite{SM}. 
In both cases, the long-time behavior of the order parameter changes 
into an exponential decay. The same happens for the density 
in the single-particle loss case.}
  
To gain further insights we derive~\cite{SM} the dynamical equation for the 
single particle density $n = \frac{1}{V}\sum_{\bk} \sigma^z_{\bk} + 1$
from Eqs.~(\ref{eq:sigmax}-\ref{eq:sigmaz})
\begin{equation}\label{eqn:dndt}
\frac{d n}{dt}=-2\Gamma\vert \Delta\vert^2-\Gamma n^2\,.
\end{equation}
The first term in Eq.~(\ref{eqn:dndt}) describes the 
depletion due to losses of Cooper pairs~\cite{yamamoto2021collective}, while the second 
one accounts for the contribution of non-condensed pairs. This term,  
\ms{which arises due to the effective single-particle losses
included in our variational approach (see Eq.~(\ref{eq:effective_lindblad}))},  
is always present even when the system is in the normal phase 
and it becomes dominant at long-times, 
correctly ensures that the steady state is the vacuum independently of the initial state.
In fact this second term is responsible for the power-law decay of the density, as one can readily 
understand by disregarding the order parameter, which gives $ \dot{n} \sim -n^2$, implying $n \sim 1/t$. 
\ms{We can now understand the long-time behavior of the 
superfluid order parameter described in Fig.~\ref{fig1}. 
Due to two-body losses, each of the $\sigma_{\mathbf{k}}^{x/y}$ 
components Anderson pseudo-spin experience
an effective single particle dissipation $\Gamma_{\rm eff}=n(t)\Gamma$
decreasing as $1/t$ at long times, leading to a $\sim 1/t $ 
power-law decay for each $\mathbf{k}-$mode. 
In addition, due to the pseudospin precession, 
each mode acquires a time dependent phase which leads 
to an additional dephasing of the order parameter.
At long times, when $|\D| \ll 1$,  
the dephasing gives and additional $\sim 1/t$ decay, thus explaining the overall $1/t^2.$
 The crossover from exponential to power-law decay in the order parameter is controlled by the time scale at which the particle density enters the power-law decay regime. On the other hand if the particle density is kept constant (by means of an additional pump) or in presence only of single particle losses, the decay rate for the pseudo-spins is constant in time,
giving rise to an order parameter which decays exponentially~\cite{SM}.}
\begin{figure}[t!]
	\includegraphics[width=\columnwidth]{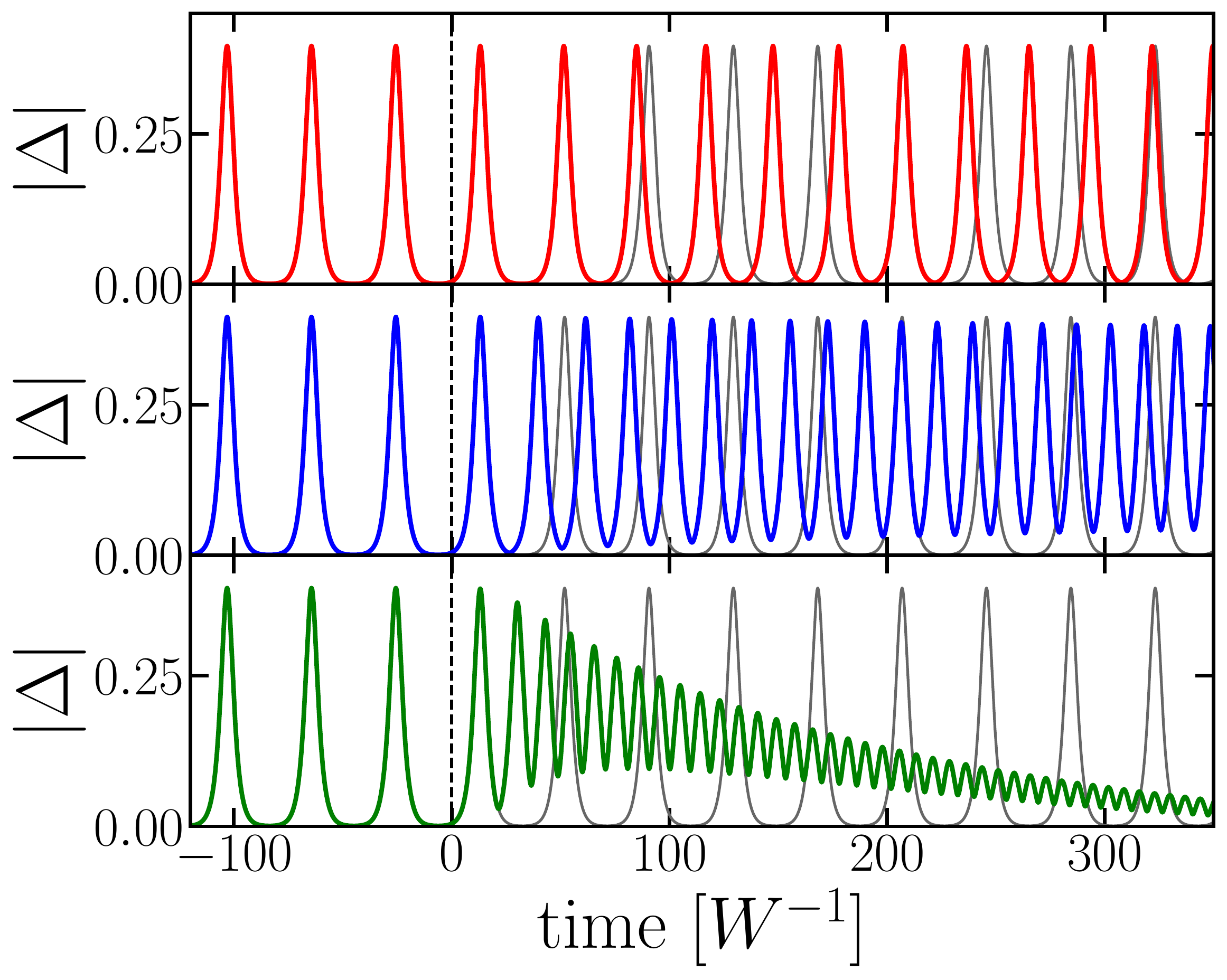}
	\caption{Dynamics after a sudden quench of the interaction $|U_i|/W = 0.125 
	\to |U_f|/W  = 1.0$. For $-200 < \mathrm{time}~W < 0 $ the dynamics is unitary.	 
	For positive times we switch on a finite dissipation, $\G/|U_f| = 10^{-7}, 10^{-4}, $ and 
	$0.005$, from top to bottom. In all the panels, 
	the light grey lines represent the corresponding unitary dynamics.}
	\label{fig2}
\end{figure}


\paragraph{Results - Double quench} We now consider the dynamics after a double quench, where first at some negative time the pairing interaction is suddenly changed $U_i\rightarrow U_f$ and then the two-body losses are suddenly switched on at time $t=0$. This dynamical protocol allows to discuss the effect of correlated dissipation on the dynamical synchronization transition~\cite{BarankovLevitovSpivakPRL04,BarankovLevitovPRL06,YuzbashyanEtAlPRB05,YuzbashyanEtAlPRL06} that is known to occur in the isolated case.

 In Fig.~\ref{fig2} we plot the dynamics of the order parameter $\Delta(t)$ after 
 a quench of the pairing attraction from $|U_i|/W=0.125$ 
 to $|U_f|/W=1.0$, 
 corresponding to the synchronized BCS regime in the isolated system, and for increasing values of two-body losses. 
 We compare this dynamics to the purely unitary case 
 (grey lines in the background of each panel) where we recognize the characteristic coherent oscillations of the order parameter, with a period controlled by the 
 ratio between initial and final gap~\cite{BarankovLevitovPRL06}. 
 We see that the switching of the dissipation at $t=0$ drastically changes the time evolution, 
 inducing not only a damping of coherent oscillations but also a substantial 
 renormalization of their frequency, which increases with $\Gamma$. 
 Remarkably we note from the upper panel of Fig.~\ref{fig2} that even a 
 tiny dissipation, corresponding to $\Gamma/\vert U_f\vert=10^{-7}$, has a sizable 
 effect on the oscillation frequency. 
 To highlight this point we extract the 
 dominant frequency $\omega_{\star}$ 
 of the coherent oscillations of the order parameter, obtained by 
 Fourier transforming the real-time 
 signal over a time window $\Delta T = 1000 W^{-1}$,
\ms{and plot the associated period $\mathcal{T}_{\star}=2\pi/\omega_{\star}$} in Fig.~\ref{fig3} as a function of $\Gamma/\vert U_f\vert$. 
  \begin{figure}[t!]
	\includegraphics[width=\columnwidth]{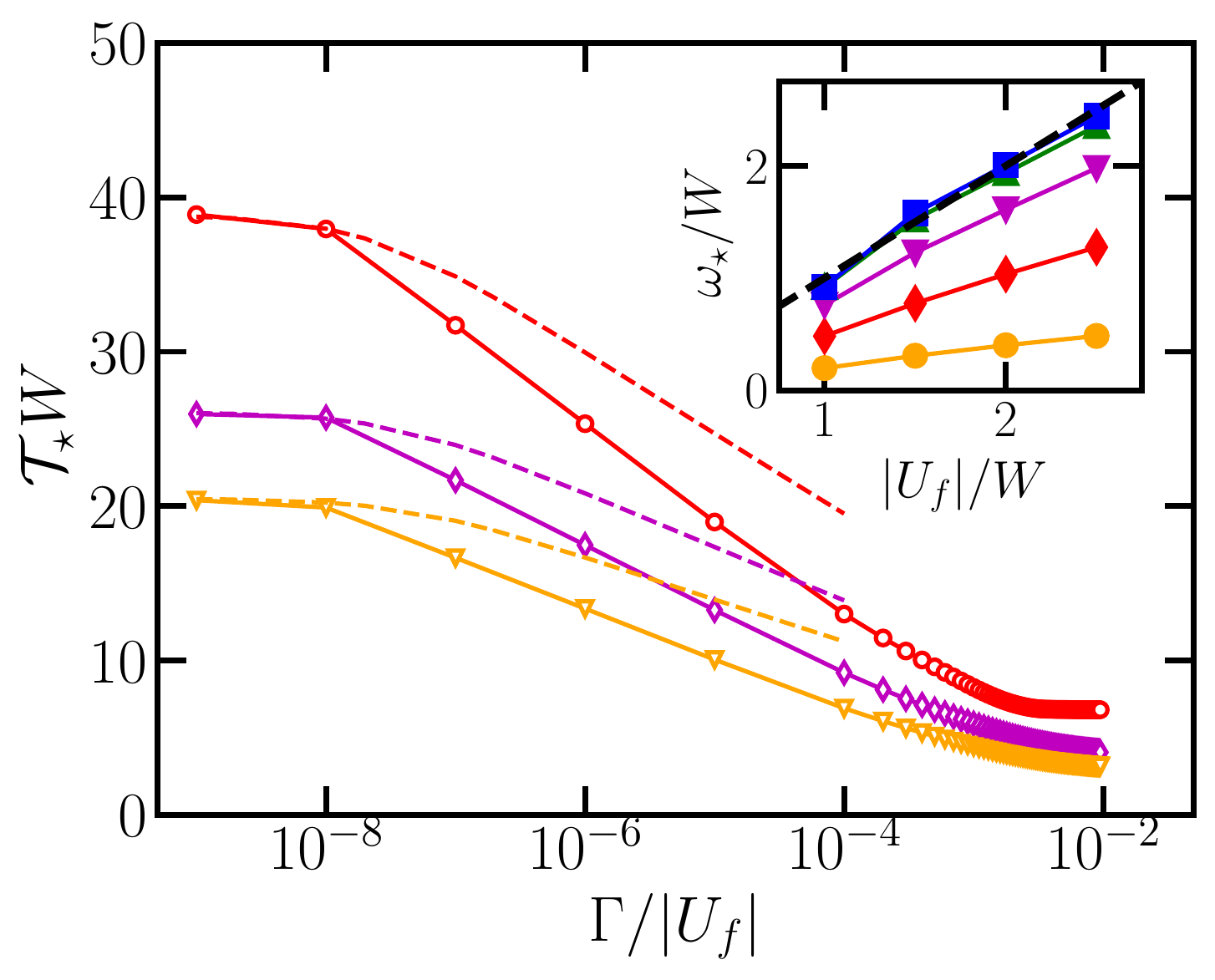}
	\caption{
	Period of coherent oscillations as a function of the dissipation
	and different values of the final interaction $|U_f |/W = 1.00$ (circles),
	$1.50$ (diamonds), and $2.00$ 
	(down triangles). 
	Dashed lines indicate the analytic estimate using the dissipative soliton solution.
	(Inset) Frequency of oscillations $\omega_{\star}=2\pi/\mathcal{T}_{\star}$ plotted as a function of $|U_f|$, and 
	different values of the $\G/|U_f| = 10^{-7}$ (circles), $10^{-4}$ (diamonds),
	$10^{-3}$ (down triangles), $5\times 10^{-3}$ (up triangles) and $10^{-2}$ (squares).
	Dashed line indicates the $\o_{\star} = |U_f|$ line.}
	\label{fig3} 
\end{figure}
\ms{
 We see that $\mathcal{T}_{\star}$ depends strongly on the losses, with a 
 non-linear behavior compatible with a logarithmic dependence 
 ${\cal T}_\star \sim - \log \left( e^{-{\cal T}_\star^0} +c \frac{\Gamma}{|U_f|} \right)$ 
 where $\mathcal{T}_\star^0$ is the oscillation period in the isolated case and $c$ a numerical prefactor.}
 As the dissipation is increased, even though remaining a small 
 fraction of the interaction $|U_f|$, 
 we see that the frequency $\o_\star$ 
 tends to saturate {to a value $\o_\star= \vert U_f\vert$, see Fig.~\ref{fig3} (inset)}. 
 
The fact that the dissipation changes so dramatically the frequency of oscillations 
of the order parameter is the second important result of this work. 
\ms{We can understand this effect by constructing a dissipative soliton solution 
for the BCS problem,  assuming that the Anderson pseudo-spin norm 
conservation is weakly broken~\cite{SM}.
For small dissipation $\Gamma/|U_f| \to 0$, we consider an effectively unitary dynamics 
with a renormalized pseudo-spin length which we determine self-consistently.
We obtain a dissipative soliton train solution~\cite{BarankovLevitovPRL06}
with period ${\cal T}_\star(\G) = 2 K(1-\a(\G)^2)/|U_f| \D_+(\G)$, 
being $\a(\G) = \D_-(\G)/\D_+(\G) $ and $K$ the complete elliptic integral 
of the first kind. The period depends on dissipation through 
the soliton amplitudes $\D_{\pm}(\G)$ obtained by the solution 
of the self-consistent equations~\cite{SM}.
Remarkably, the dissipative soliton captures 
qualitatively well the logarithmic dependence of 
the oscillation period  on the 
dissipation, see Fig.~\ref{fig3} dashed lines.
Quantitatively, the dissipative solitons 
underestimate the frequency renormalization.
This can be expected as the above argument 
is strictly valid only for times $t \sim {\cal T}_\star$, 
whereas the numerical frequencies 
are extracted by averaging over a large number of periods $\D T \gg {\cal T}_\star$.
At large times, $\G t \gg 1$ the dissipative solitons 
are washed away by the decay of the non-equilibrium superconducting pairs 
which occurs through excitations of energy $\sim |U_f| \gg 2 \pi / {\cal T}_\star$
and determine a faster oscillatory dynamics.
For $\G/|U_f| \gtrsim 10^{-2}$ the decay dynamics 
quickly takes over the dissipative solitons, 
thus leading to the observed saturation to
$\omega_{\star} = |U_f|$.}

\paragraph{Conclusions - } In this work we have studied the dissipative dynamics of a 
fermionic superfluid with two-body losses. We have used a time-dependent variational 
method for open quantum systems, from which the resulting  dynamics takes the form a 
BCS problem with complex pairing interactions and effective single particle losses
that were disregarded in previous works~\cite{yamamoto2021collective,YamamotoEtAlPRL19}. 
\ms{We show that the latter plays a key role for the dynamics of the system. It underlies both the power-law
decay of particle density and order parameter after a dissipation quench as well as the strong
renormalization of the period of coherent oscillations after a quench of dissipation and pairing interaction.
We have qualitavely captured this latter effect in terms of a dissipative soliton solution.}
%

\ms{Our results highlight the non trivial nature of many-body dissipative processes in giving rise to universal dynamical regimes and, in particular, to novel regimes of superfluidity. Future directions opened by this work include the study of Zeno-like superfluid dynamics in the strongly dissipative regime as well as the application of the variational dynamics to strongly correlated dissipative many-body systems, such as the dissipative Fermi-Hubbard model~\cite{honda2022observation}. Finally, it would be interesting to investigate how the signatures of this dissipative dynamics can be observed in experiments with dissipative gases and cavity QED simulators.}

\paragraph{Acknowledgments} We thank A. Chiocchetta for collaboration at an early stage of the project. MS acknowledges support by the ANR grant "NonEQuMat" (ANR-19-CE47-0001) and from the European Research Council (ERC) under the Eu-
ropean Union’s Horizon 2020 research and innovation programme (Grant agreement No. 101002955 — CONQUER).
G.M. acknowledges  support by the Swiss National Science Foundation through an AMBIZIONE grant, and by the MUR - Italian Minister of University and Research through  the ``Rita Levi-Montalcini" program.


\begin{thebibliography}{80}%
\makeatletter
\providecommand \@ifxundefined [1]{%
 \@ifx{#1\undefined}
}%
\providecommand \@ifnum [1]{%
 \ifnum #1\expandafter \@firstoftwo
 \else \expandafter \@secondoftwo
 \fi
}%
\providecommand \@ifx [1]{%
 \ifx #1\expandafter \@firstoftwo
 \else \expandafter \@secondoftwo
 \fi
}%
\providecommand \natexlab [1]{#1}%
\providecommand \enquote  [1]{``#1''}%
\providecommand \bibnamefont  [1]{#1}%
\providecommand \bibfnamefont [1]{#1}%
\providecommand \citenamefont [1]{#1}%
\providecommand \href@noop [0]{\@secondoftwo}%
\providecommand \href [0]{\begingroup \@sanitize@url \@href}%
\providecommand \@href[1]{\@@startlink{#1}\@@href}%
\providecommand \@@href[1]{\endgroup#1\@@endlink}%
\providecommand \@sanitize@url [0]{\catcode `\\12\catcode `\$12\catcode
  `\&12\catcode `\#12\catcode `\^12\catcode `\_12\catcode `\%12\relax}%
\providecommand \@@startlink[1]{}%
\providecommand \@@endlink[0]{}%
\providecommand \url  [0]{\begingroup\@sanitize@url \@url }%
\providecommand \@url [1]{\endgroup\@href {#1}{\urlprefix }}%
\providecommand \urlprefix  [0]{URL }%
\providecommand \Eprint [0]{\href }%
\providecommand \doibase [0]{http://dx.doi.org/}%
\providecommand \selectlanguage [0]{\@gobble}%
\providecommand \bibinfo  [0]{\@secondoftwo}%
\providecommand \bibfield  [0]{\@secondoftwo}%
\providecommand \translation [1]{[#1]}%
\providecommand \BibitemOpen [0]{}%
\providecommand \bibitemStop [0]{}%
\providecommand \bibitemNoStop [0]{.\EOS\space}%
\providecommand \EOS [0]{\spacefactor3000\relax}%
\providecommand \BibitemShut  [1]{\csname bibitem#1\endcsname}%
\let\auto@bib@innerbib\@empty
\bibitem [{\citenamefont {Matsunaga}\ \emph {et~al.}(2014)\citenamefont
  {Matsunaga}, \citenamefont {Tsuji}, \citenamefont {Fujita}, \citenamefont
  {Sugioka}, \citenamefont {Makise}, \citenamefont {Uzawa}, \citenamefont
  {Terai}, \citenamefont {Wang}, \citenamefont {Aoki},\ and\ \citenamefont
  {Shimano}}]{matsunaga2014light}%
  \BibitemOpen
  \bibfield  {author} {\bibinfo {author} {\bibfnamefont {R.}~\bibnamefont
  {Matsunaga}}, \bibinfo {author} {\bibfnamefont {N.}~\bibnamefont {Tsuji}},
  \bibinfo {author} {\bibfnamefont {H.}~\bibnamefont {Fujita}}, \bibinfo
  {author} {\bibfnamefont {A.}~\bibnamefont {Sugioka}}, \bibinfo {author}
  {\bibfnamefont {K.}~\bibnamefont {Makise}}, \bibinfo {author} {\bibfnamefont
  {Y.}~\bibnamefont {Uzawa}}, \bibinfo {author} {\bibfnamefont
  {H.}~\bibnamefont {Terai}}, \bibinfo {author} {\bibfnamefont
  {Z.}~\bibnamefont {Wang}}, \bibinfo {author} {\bibfnamefont {H.}~\bibnamefont
  {Aoki}}, \ and\ \bibinfo {author} {\bibfnamefont {R.}~\bibnamefont
  {Shimano}},\ }\href {\doibase 10.1126/science.1254697} {\bibfield  {journal}
  {\bibinfo  {journal} {Science}\ }\textbf {\bibinfo {volume} {345}},\ \bibinfo
  {pages} {1145} (\bibinfo {year} {2014})},\ \Eprint
  {http://arxiv.org/abs/https://www.science.org/doi/pdf/10.1126/science.1254697}
  {https://www.science.org/doi/pdf/10.1126/science.1254697} \BibitemShut
  {NoStop}%
\bibitem [{\citenamefont {Shimano}\ and\ \citenamefont
  {Tsuji}(2020)}]{shimano2020higgs}%
  \BibitemOpen
  \bibfield  {author} {\bibinfo {author} {\bibfnamefont {R.}~\bibnamefont
  {Shimano}}\ and\ \bibinfo {author} {\bibfnamefont {N.}~\bibnamefont
  {Tsuji}},\ }\href {\doibase 10.1146/annurev-conmatphys-031119-050813}
  {\bibfield  {journal} {\bibinfo  {journal} {Annual Review of Condensed Matter
  Physics}\ }\textbf {\bibinfo {volume} {11}},\ \bibinfo {pages} {103}
  (\bibinfo {year} {2020})},\ \Eprint
  {http://arxiv.org/abs/https://doi.org/10.1146/annurev-conmatphys-031119-050813}
  {https://doi.org/10.1146/annurev-conmatphys-031119-050813} \BibitemShut
  {NoStop}%
\bibitem [{\citenamefont {Fausti}\ \emph {et~al.}(2011)\citenamefont {Fausti},
  \citenamefont {Tobey}, \citenamefont {Dean}, \citenamefont {Kaiser},
  \citenamefont {Dienst}, \citenamefont {Hoffmann}, \citenamefont {Pyon},
  \citenamefont {Takayama}, \citenamefont {Takagi},\ and\ \citenamefont
  {Cavalleri}}]{Fausti_Science11}%
  \BibitemOpen
  \bibfield  {author} {\bibinfo {author} {\bibfnamefont {D.}~\bibnamefont
  {Fausti}}, \bibinfo {author} {\bibfnamefont {R.~I.}\ \bibnamefont {Tobey}},
  \bibinfo {author} {\bibfnamefont {N.}~\bibnamefont {Dean}}, \bibinfo {author}
  {\bibfnamefont {S.}~\bibnamefont {Kaiser}}, \bibinfo {author} {\bibfnamefont
  {A.}~\bibnamefont {Dienst}}, \bibinfo {author} {\bibfnamefont {M.~C.}\
  \bibnamefont {Hoffmann}}, \bibinfo {author} {\bibfnamefont {S.}~\bibnamefont
  {Pyon}}, \bibinfo {author} {\bibfnamefont {T.}~\bibnamefont {Takayama}},
  \bibinfo {author} {\bibfnamefont {H.}~\bibnamefont {Takagi}}, \ and\ \bibinfo
  {author} {\bibfnamefont {A.}~\bibnamefont {Cavalleri}},\ }\href {\doibase
  10.1126/science.1197294} {\bibfield  {journal} {\bibinfo  {journal}
  {Science}\ }\textbf {\bibinfo {volume} {331}},\ \bibinfo {pages} {189}
  (\bibinfo {year} {2011})}\BibitemShut {NoStop}%
\bibitem [{\citenamefont {Mitrano}\ \emph {et~al.}(2016)\citenamefont
  {Mitrano}, \citenamefont {Cantaluppi}, \citenamefont {Nicoletti},
  \citenamefont {Kaiser}, \citenamefont {Perucchi}, \citenamefont {Lupi},
  \citenamefont {Di~Pietro}, \citenamefont {Pontiroli}, \citenamefont
  {Ricc{\`o}}, \citenamefont {Clark}, \citenamefont {Jaksch},\ and\
  \citenamefont {Cavalleri}}]{MitranoNature16}%
  \BibitemOpen
  \bibfield  {author} {\bibinfo {author} {\bibfnamefont {M.}~\bibnamefont
  {Mitrano}}, \bibinfo {author} {\bibfnamefont {A.}~\bibnamefont {Cantaluppi}},
  \bibinfo {author} {\bibfnamefont {D.}~\bibnamefont {Nicoletti}}, \bibinfo
  {author} {\bibfnamefont {S.}~\bibnamefont {Kaiser}}, \bibinfo {author}
  {\bibfnamefont {A.}~\bibnamefont {Perucchi}}, \bibinfo {author}
  {\bibfnamefont {S.}~\bibnamefont {Lupi}}, \bibinfo {author} {\bibfnamefont
  {P.}~\bibnamefont {Di~Pietro}}, \bibinfo {author} {\bibfnamefont
  {D.}~\bibnamefont {Pontiroli}}, \bibinfo {author} {\bibfnamefont
  {M.}~\bibnamefont {Ricc{\`o}}}, \bibinfo {author} {\bibfnamefont {S.~R.}\
  \bibnamefont {Clark}}, \bibinfo {author} {\bibfnamefont {D.}~\bibnamefont
  {Jaksch}}, \ and\ \bibinfo {author} {\bibfnamefont {A.}~\bibnamefont
  {Cavalleri}},\ }\href@noop {} {\bibfield  {journal} {\bibinfo  {journal}
  {Nature}\ }\textbf {\bibinfo {volume} {530}},\ \bibinfo {pages} {461}
  (\bibinfo {year} {2016})}\BibitemShut {NoStop}%
\bibitem [{\citenamefont {Buzzi}\ \emph {et~al.}(2020)\citenamefont {Buzzi},
  \citenamefont {Nicoletti}, \citenamefont {Fechner}, \citenamefont
  {Tancogne-Dejean}, \citenamefont {Sentef}, \citenamefont {Georges},
  \citenamefont {Biesner}, \citenamefont {Uykur}, \citenamefont {Dressel},
  \citenamefont {Henderson}, \citenamefont {Siegrist}, \citenamefont
  {Schlueter}, \citenamefont {Miyagawa}, \citenamefont {Kanoda}, \citenamefont
  {Nam}, \citenamefont {Ardavan}, \citenamefont {Coulthard}, \citenamefont
  {Tindall}, \citenamefont {Schlawin}, \citenamefont {Jaksch},\ and\
  \citenamefont {Cavalleri}}]{Buzzi2020photomolecular}%
  \BibitemOpen
  \bibfield  {author} {\bibinfo {author} {\bibfnamefont {M.}~\bibnamefont
  {Buzzi}}, \bibinfo {author} {\bibfnamefont {D.}~\bibnamefont {Nicoletti}},
  \bibinfo {author} {\bibfnamefont {M.}~\bibnamefont {Fechner}}, \bibinfo
  {author} {\bibfnamefont {N.}~\bibnamefont {Tancogne-Dejean}}, \bibinfo
  {author} {\bibfnamefont {M.~A.}\ \bibnamefont {Sentef}}, \bibinfo {author}
  {\bibfnamefont {A.}~\bibnamefont {Georges}}, \bibinfo {author} {\bibfnamefont
  {T.}~\bibnamefont {Biesner}}, \bibinfo {author} {\bibfnamefont
  {E.}~\bibnamefont {Uykur}}, \bibinfo {author} {\bibfnamefont
  {M.}~\bibnamefont {Dressel}}, \bibinfo {author} {\bibfnamefont
  {A.}~\bibnamefont {Henderson}}, \bibinfo {author} {\bibfnamefont
  {T.}~\bibnamefont {Siegrist}}, \bibinfo {author} {\bibfnamefont {J.~A.}\
  \bibnamefont {Schlueter}}, \bibinfo {author} {\bibfnamefont {K.}~\bibnamefont
  {Miyagawa}}, \bibinfo {author} {\bibfnamefont {K.}~\bibnamefont {Kanoda}},
  \bibinfo {author} {\bibfnamefont {M.-S.}\ \bibnamefont {Nam}}, \bibinfo
  {author} {\bibfnamefont {A.}~\bibnamefont {Ardavan}}, \bibinfo {author}
  {\bibfnamefont {J.}~\bibnamefont {Coulthard}}, \bibinfo {author}
  {\bibfnamefont {J.}~\bibnamefont {Tindall}}, \bibinfo {author} {\bibfnamefont
  {F.}~\bibnamefont {Schlawin}}, \bibinfo {author} {\bibfnamefont
  {D.}~\bibnamefont {Jaksch}}, \ and\ \bibinfo {author} {\bibfnamefont
  {A.}~\bibnamefont {Cavalleri}},\ }\href {\doibase 10.1103/PhysRevX.10.031028}
  {\bibfield  {journal} {\bibinfo  {journal} {Phys. Rev. X}\ }\textbf {\bibinfo
  {volume} {10}},\ \bibinfo {pages} {031028} (\bibinfo {year}
  {2020})}\BibitemShut {NoStop}%
\bibitem [{\citenamefont {Budden}\ \emph {et~al.}(2021)\citenamefont {Budden},
  \citenamefont {Gebert}, \citenamefont {Buzzi}, \citenamefont {Jotzu},
  \citenamefont {Wang}, \citenamefont {Matsuyama}, \citenamefont {Meier},
  \citenamefont {Laplace}, \citenamefont {Pontiroli}, \citenamefont
  {Ricc{\`o}}, \citenamefont {Schlawin}, \citenamefont {Jaksch},\ and\
  \citenamefont {Cavalleri}}]{budden2021evidence}%
  \BibitemOpen
  \bibfield  {author} {\bibinfo {author} {\bibfnamefont {M.}~\bibnamefont
  {Budden}}, \bibinfo {author} {\bibfnamefont {T.}~\bibnamefont {Gebert}},
  \bibinfo {author} {\bibfnamefont {M.}~\bibnamefont {Buzzi}}, \bibinfo
  {author} {\bibfnamefont {G.}~\bibnamefont {Jotzu}}, \bibinfo {author}
  {\bibfnamefont {E.}~\bibnamefont {Wang}}, \bibinfo {author} {\bibfnamefont
  {T.}~\bibnamefont {Matsuyama}}, \bibinfo {author} {\bibfnamefont
  {G.}~\bibnamefont {Meier}}, \bibinfo {author} {\bibfnamefont
  {Y.}~\bibnamefont {Laplace}}, \bibinfo {author} {\bibfnamefont
  {D.}~\bibnamefont {Pontiroli}}, \bibinfo {author} {\bibfnamefont
  {M.}~\bibnamefont {Ricc{\`o}}}, \bibinfo {author} {\bibfnamefont
  {F.}~\bibnamefont {Schlawin}}, \bibinfo {author} {\bibfnamefont
  {D.}~\bibnamefont {Jaksch}}, \ and\ \bibinfo {author} {\bibfnamefont
  {A.}~\bibnamefont {Cavalleri}},\ }\href {\doibase 10.1038/s41567-020-01148-1}
  {\bibfield  {journal} {\bibinfo  {journal} {Nature Physics}\ }\textbf
  {\bibinfo {volume} {17}},\ \bibinfo {pages} {611} (\bibinfo {year}
  {2021})}\BibitemShut {NoStop}%
\bibitem [{\citenamefont {Giannetti}\ \emph {et~al.}(2016)\citenamefont
  {Giannetti}, \citenamefont {Capone}, \citenamefont {Fausti}, \citenamefont
  {Fabrizio}, \citenamefont {Parmigiani},\ and\ \citenamefont
  {Mihailovic}}]{giannetti2016ultrafast}%
  \BibitemOpen
  \bibfield  {author} {\bibinfo {author} {\bibfnamefont {C.}~\bibnamefont
  {Giannetti}}, \bibinfo {author} {\bibfnamefont {M.}~\bibnamefont {Capone}},
  \bibinfo {author} {\bibfnamefont {D.}~\bibnamefont {Fausti}}, \bibinfo
  {author} {\bibfnamefont {M.}~\bibnamefont {Fabrizio}}, \bibinfo {author}
  {\bibfnamefont {F.}~\bibnamefont {Parmigiani}}, \ and\ \bibinfo {author}
  {\bibfnamefont {D.}~\bibnamefont {Mihailovic}},\ }\href {\doibase
  10.1080/00018732.2016.1194044} {\bibfield  {journal} {\bibinfo  {journal}
  {Advances in Physics}\ }\textbf {\bibinfo {volume} {65}},\ \bibinfo {pages}
  {58} (\bibinfo {year} {2016})},\ \Eprint
  {http://arxiv.org/abs/https://doi.org/10.1080/00018732.2016.1194044}
  {https://doi.org/10.1080/00018732.2016.1194044} \BibitemShut {NoStop}%
\bibitem [{\citenamefont {Oka}\ and\ \citenamefont {Kitamura}(2019)}]{Oka2018}%
  \BibitemOpen
  \bibfield  {author} {\bibinfo {author} {\bibfnamefont {T.}~\bibnamefont
  {Oka}}\ and\ \bibinfo {author} {\bibfnamefont {S.}~\bibnamefont {Kitamura}},\
  }\href
  {https://www.annualreviews.org/doi/10.1146/annurev-conmatphys-031218-013423
  http://arxiv.org/abs/1804.03212} {\bibfield  {journal} {\bibinfo  {journal}
  {Annu. Rev.\ Condens. Matter Phys.}\ }\textbf {\bibinfo {volume} {10}},\
  \bibinfo {pages} {387} (\bibinfo {year} {2019})},\ \Eprint
  {http://arxiv.org/abs/1804.03212} {arXiv:1804.03212} \BibitemShut {NoStop}%
\bibitem [{\citenamefont {de~la Torre}\ \emph {et~al.}(2021)\citenamefont
  {de~la Torre}, \citenamefont {Kennes}, \citenamefont {Claassen},
  \citenamefont {Gerber}, \citenamefont {McIver},\ and\ \citenamefont
  {Sentef}}]{delatorre2021nonthermal}%
  \BibitemOpen
  \bibfield  {author} {\bibinfo {author} {\bibfnamefont {A.}~\bibnamefont
  {de~la Torre}}, \bibinfo {author} {\bibfnamefont {D.~M.}\ \bibnamefont
  {Kennes}}, \bibinfo {author} {\bibfnamefont {M.}~\bibnamefont {Claassen}},
  \bibinfo {author} {\bibfnamefont {S.}~\bibnamefont {Gerber}}, \bibinfo
  {author} {\bibfnamefont {J.~W.}\ \bibnamefont {McIver}}, \ and\ \bibinfo
  {author} {\bibfnamefont {M.~A.}\ \bibnamefont {Sentef}},\ }\href {\doibase
  10.1103/RevModPhys.93.041002} {\bibfield  {journal} {\bibinfo  {journal}
  {Rev. Mod. Phys.}\ }\textbf {\bibinfo {volume} {93}},\ \bibinfo {pages}
  {041002} (\bibinfo {year} {2021})}\BibitemShut {NoStop}%
\bibitem [{\citenamefont {Regal}\ \emph {et~al.}(2004)\citenamefont {Regal},
  \citenamefont {Greiner},\ and\ \citenamefont {Jin}}]{regal2004observation}%
  \BibitemOpen
  \bibfield  {author} {\bibinfo {author} {\bibfnamefont {C.~A.}\ \bibnamefont
  {Regal}}, \bibinfo {author} {\bibfnamefont {M.}~\bibnamefont {Greiner}}, \
  and\ \bibinfo {author} {\bibfnamefont {D.~S.}\ \bibnamefont {Jin}},\ }\href
  {\doibase 10.1103/PhysRevLett.92.040403} {\bibfield  {journal} {\bibinfo
  {journal} {Phys. Rev. Lett.}\ }\textbf {\bibinfo {volume} {92}},\ \bibinfo
  {pages} {040403} (\bibinfo {year} {2004})}\BibitemShut {NoStop}%
\bibitem [{\citenamefont {Zwierlein}\ \emph {et~al.}(2004)\citenamefont
  {Zwierlein}, \citenamefont {Stan}, \citenamefont {Schunck}, \citenamefont
  {Raupach}, \citenamefont {Kerman},\ and\ \citenamefont
  {Ketterle}}]{zwierlein2004condensation}%
  \BibitemOpen
  \bibfield  {author} {\bibinfo {author} {\bibfnamefont {M.~W.}\ \bibnamefont
  {Zwierlein}}, \bibinfo {author} {\bibfnamefont {C.~A.}\ \bibnamefont {Stan}},
  \bibinfo {author} {\bibfnamefont {C.~H.}\ \bibnamefont {Schunck}}, \bibinfo
  {author} {\bibfnamefont {S.~M.~F.}\ \bibnamefont {Raupach}}, \bibinfo
  {author} {\bibfnamefont {A.~J.}\ \bibnamefont {Kerman}}, \ and\ \bibinfo
  {author} {\bibfnamefont {W.}~\bibnamefont {Ketterle}},\ }\href {\doibase
  10.1103/PhysRevLett.92.120403} {\bibfield  {journal} {\bibinfo  {journal}
  {Phys. Rev. Lett.}\ }\textbf {\bibinfo {volume} {92}},\ \bibinfo {pages}
  {120403} (\bibinfo {year} {2004})}\BibitemShut {NoStop}%
\bibitem [{\citenamefont {Bartenstein}\ \emph {et~al.}(2004)\citenamefont
  {Bartenstein}, \citenamefont {Altmeyer}, \citenamefont {Riedl}, \citenamefont
  {Jochim}, \citenamefont {Chin}, \citenamefont {Denschlag},\ and\
  \citenamefont {Grimm}}]{bartenstein2004crossover}%
  \BibitemOpen
  \bibfield  {author} {\bibinfo {author} {\bibfnamefont {M.}~\bibnamefont
  {Bartenstein}}, \bibinfo {author} {\bibfnamefont {A.}~\bibnamefont
  {Altmeyer}}, \bibinfo {author} {\bibfnamefont {S.}~\bibnamefont {Riedl}},
  \bibinfo {author} {\bibfnamefont {S.}~\bibnamefont {Jochim}}, \bibinfo
  {author} {\bibfnamefont {C.}~\bibnamefont {Chin}}, \bibinfo {author}
  {\bibfnamefont {J.~H.}\ \bibnamefont {Denschlag}}, \ and\ \bibinfo {author}
  {\bibfnamefont {R.}~\bibnamefont {Grimm}},\ }\href {\doibase
  10.1103/PhysRevLett.92.120401} {\bibfield  {journal} {\bibinfo  {journal}
  {Phys. Rev. Lett.}\ }\textbf {\bibinfo {volume} {92}},\ \bibinfo {pages}
  {120401} (\bibinfo {year} {2004})}\BibitemShut {NoStop}%
\bibitem [{\citenamefont {Behrle}\ \emph {et~al.}(2018)\citenamefont {Behrle},
  \citenamefont {Harrison}, \citenamefont {Kombe}, \citenamefont {Gao},
  \citenamefont {Link}, \citenamefont {Bernier}, \citenamefont {Kollath},\ and\
  \citenamefont {K{\"o}hl}}]{behrle2019higgs}%
  \BibitemOpen
  \bibfield  {author} {\bibinfo {author} {\bibfnamefont {A.}~\bibnamefont
  {Behrle}}, \bibinfo {author} {\bibfnamefont {T.}~\bibnamefont {Harrison}},
  \bibinfo {author} {\bibfnamefont {J.}~\bibnamefont {Kombe}}, \bibinfo
  {author} {\bibfnamefont {K.}~\bibnamefont {Gao}}, \bibinfo {author}
  {\bibfnamefont {M.}~\bibnamefont {Link}}, \bibinfo {author} {\bibfnamefont
  {J.~S.}\ \bibnamefont {Bernier}}, \bibinfo {author} {\bibfnamefont
  {C.}~\bibnamefont {Kollath}}, \ and\ \bibinfo {author} {\bibfnamefont
  {M.}~\bibnamefont {K{\"o}hl}},\ }\href {\doibase 10.1038/s41567-018-0128-6}
  {\bibfield  {journal} {\bibinfo  {journal} {Nature Physics}\ }\textbf
  {\bibinfo {volume} {14}},\ \bibinfo {pages} {781} (\bibinfo {year}
  {2018})}\BibitemShut {NoStop}%
\bibitem [{\citenamefont {Barankov}\ \emph {et~al.}(2004)\citenamefont
  {Barankov}, \citenamefont {Levitov},\ and\ \citenamefont
  {Spivak}}]{BarankovLevitovSpivakPRL04}%
  \BibitemOpen
  \bibfield  {author} {\bibinfo {author} {\bibfnamefont {R.~A.}\ \bibnamefont
  {Barankov}}, \bibinfo {author} {\bibfnamefont {L.~S.}\ \bibnamefont
  {Levitov}}, \ and\ \bibinfo {author} {\bibfnamefont {B.~Z.}\ \bibnamefont
  {Spivak}},\ }\href@noop {} {\bibfield  {journal} {\bibinfo  {journal} {Phys.
  Rev. Lett.}\ }\textbf {\bibinfo {volume} {93}},\ \bibinfo {pages} {160401}
  (\bibinfo {year} {2004})}\BibitemShut {NoStop}%
\bibitem [{\citenamefont {Barankov}\ and\ \citenamefont
  {Levitov}(2006)}]{BarankovLevitovPRL06}%
  \BibitemOpen
  \bibfield  {author} {\bibinfo {author} {\bibfnamefont {R.~A.}\ \bibnamefont
  {Barankov}}\ and\ \bibinfo {author} {\bibfnamefont {L.~S.}\ \bibnamefont
  {Levitov}},\ }\href@noop {} {\bibfield  {journal} {\bibinfo  {journal} {Phys.
  Rev. Lett.}\ }\textbf {\bibinfo {volume} {96}},\ \bibinfo {pages} {230403}
  (\bibinfo {year} {2006})}\BibitemShut {NoStop}%
\bibitem [{\citenamefont {Yuzbashyan}\ \emph {et~al.}(2005)\citenamefont
  {Yuzbashyan}, \citenamefont {Altshuler}, \citenamefont {Kuznetsov},\ and\
  \citenamefont {Enolskii}}]{YuzbashyanEtAlPRB05}%
  \BibitemOpen
  \bibfield  {author} {\bibinfo {author} {\bibfnamefont {E.~A.}\ \bibnamefont
  {Yuzbashyan}}, \bibinfo {author} {\bibfnamefont {B.~L.}\ \bibnamefont
  {Altshuler}}, \bibinfo {author} {\bibfnamefont {V.~B.}\ \bibnamefont
  {Kuznetsov}}, \ and\ \bibinfo {author} {\bibfnamefont {V.~Z.}\ \bibnamefont
  {Enolskii}},\ }\href@noop {} {\bibfield  {journal} {\bibinfo  {journal}
  {Phys. Rev. B}\ }\textbf {\bibinfo {volume} {72}},\ \bibinfo {pages} {220503}
  (\bibinfo {year} {2005})}\BibitemShut {NoStop}%
\bibitem [{\citenamefont {Yuzbashyan}\ \emph {et~al.}(2006)\citenamefont
  {Yuzbashyan}, \citenamefont {Tsyplyatyev},\ and\ \citenamefont
  {Altshuler}}]{YuzbashyanEtAlPRL06}%
  \BibitemOpen
  \bibfield  {author} {\bibinfo {author} {\bibfnamefont {E.~A.}\ \bibnamefont
  {Yuzbashyan}}, \bibinfo {author} {\bibfnamefont {O.}~\bibnamefont
  {Tsyplyatyev}}, \ and\ \bibinfo {author} {\bibfnamefont {B.~L.}\ \bibnamefont
  {Altshuler}},\ }\href@noop {} {\bibfield  {journal} {\bibinfo  {journal}
  {Phys. Rev. Lett.}\ }\textbf {\bibinfo {volume} {96}},\ \bibinfo {pages}
  {097005} (\bibinfo {year} {2006})}\BibitemShut {NoStop}%
\bibitem [{\citenamefont {Gurarie}(2009)}]{GurariePRL09}%
  \BibitemOpen
  \bibfield  {author} {\bibinfo {author} {\bibfnamefont {V.}~\bibnamefont
  {Gurarie}},\ }\href@noop {} {\bibfield  {journal} {\bibinfo  {journal} {Phys.
  Rev. Lett.}\ }\textbf {\bibinfo {volume} {103}},\ \bibinfo {pages} {075301}
  (\bibinfo {year} {2009})}\BibitemShut {NoStop}%
\bibitem [{\citenamefont {Foster}\ \emph {et~al.}(2014)\citenamefont {Foster},
  \citenamefont {Gurarie}, \citenamefont {Dzero},\ and\ \citenamefont
  {Yuzbashyan}}]{foster2014quench}%
  \BibitemOpen
  \bibfield  {author} {\bibinfo {author} {\bibfnamefont {M.~S.}\ \bibnamefont
  {Foster}}, \bibinfo {author} {\bibfnamefont {V.}~\bibnamefont {Gurarie}},
  \bibinfo {author} {\bibfnamefont {M.}~\bibnamefont {Dzero}}, \ and\ \bibinfo
  {author} {\bibfnamefont {E.~A.}\ \bibnamefont {Yuzbashyan}},\ }\href
  {\doibase 10.1103/PhysRevLett.113.076403} {\bibfield  {journal} {\bibinfo
  {journal} {Phys. Rev. Lett.}\ }\textbf {\bibinfo {volume} {113}},\ \bibinfo
  {pages} {076403} (\bibinfo {year} {2014})}\BibitemShut {NoStop}%
\bibitem [{\citenamefont {Kurkjian}\ \emph {et~al.}(2019)\citenamefont
  {Kurkjian}, \citenamefont {Klimin}, \citenamefont {Tempere},\ and\
  \citenamefont {Castin}}]{kurkjian2019pair}%
  \BibitemOpen
  \bibfield  {author} {\bibinfo {author} {\bibfnamefont {H.}~\bibnamefont
  {Kurkjian}}, \bibinfo {author} {\bibfnamefont {S.~N.}\ \bibnamefont
  {Klimin}}, \bibinfo {author} {\bibfnamefont {J.}~\bibnamefont {Tempere}}, \
  and\ \bibinfo {author} {\bibfnamefont {Y.}~\bibnamefont {Castin}},\ }\href
  {\doibase 10.1103/PhysRevLett.122.093403} {\bibfield  {journal} {\bibinfo
  {journal} {Phys. Rev. Lett.}\ }\textbf {\bibinfo {volume} {122}},\ \bibinfo
  {pages} {093403} (\bibinfo {year} {2019})}\BibitemShut {NoStop}%
\bibitem [{\citenamefont {Mazza}\ and\ \citenamefont
  {Georges}(2017)}]{mazza2017nonequilibrium}%
  \BibitemOpen
  \bibfield  {author} {\bibinfo {author} {\bibfnamefont {G.}~\bibnamefont
  {Mazza}}\ and\ \bibinfo {author} {\bibfnamefont {A.}~\bibnamefont
  {Georges}},\ }\href {\doibase 10.1103/PhysRevB.96.064515} {\bibfield
  {journal} {\bibinfo  {journal} {Phys. Rev. B}\ }\textbf {\bibinfo {volume}
  {96}},\ \bibinfo {pages} {064515} (\bibinfo {year} {2017})}\BibitemShut
  {NoStop}%
\bibitem [{\citenamefont {Babadi}\ \emph {et~al.}(2017)\citenamefont {Babadi},
  \citenamefont {Knap}, \citenamefont {Martin}, \citenamefont {Refael},\ and\
  \citenamefont {Demler}}]{babadi2017theory}%
  \BibitemOpen
  \bibfield  {author} {\bibinfo {author} {\bibfnamefont {M.}~\bibnamefont
  {Babadi}}, \bibinfo {author} {\bibfnamefont {M.}~\bibnamefont {Knap}},
  \bibinfo {author} {\bibfnamefont {I.}~\bibnamefont {Martin}}, \bibinfo
  {author} {\bibfnamefont {G.}~\bibnamefont {Refael}}, \ and\ \bibinfo {author}
  {\bibfnamefont {E.}~\bibnamefont {Demler}},\ }\href {\doibase
  10.1103/PhysRevB.96.014512} {\bibfield  {journal} {\bibinfo  {journal} {Phys.
  Rev. B}\ }\textbf {\bibinfo {volume} {96}},\ \bibinfo {pages} {014512}
  (\bibinfo {year} {2017})}\BibitemShut {NoStop}%
\bibitem [{\citenamefont {Nava}\ \emph {et~al.}(2018)\citenamefont {Nava},
  \citenamefont {Giannetti}, \citenamefont {Georges}, \citenamefont {Tosatti},\
  and\ \citenamefont {Fabrizio}}]{nava2018cooling}%
  \BibitemOpen
  \bibfield  {author} {\bibinfo {author} {\bibfnamefont {A.}~\bibnamefont
  {Nava}}, \bibinfo {author} {\bibfnamefont {C.}~\bibnamefont {Giannetti}},
  \bibinfo {author} {\bibfnamefont {A.}~\bibnamefont {Georges}}, \bibinfo
  {author} {\bibfnamefont {E.}~\bibnamefont {Tosatti}}, \ and\ \bibinfo
  {author} {\bibfnamefont {M.}~\bibnamefont {Fabrizio}},\ }\href {\doibase
  10.1038/nphys4288} {\bibfield  {journal} {\bibinfo  {journal} {Nature
  Physics}\ }\textbf {\bibinfo {volume} {14}},\ \bibinfo {pages} {154}
  (\bibinfo {year} {2018})}\BibitemShut {NoStop}%
\bibitem [{\citenamefont {Li}\ \emph {et~al.}(2020)\citenamefont {Li},
  \citenamefont {Golez}, \citenamefont {Werner},\ and\ \citenamefont
  {Eckstein}}]{li2020eta}%
  \BibitemOpen
  \bibfield  {author} {\bibinfo {author} {\bibfnamefont {J.}~\bibnamefont
  {Li}}, \bibinfo {author} {\bibfnamefont {D.}~\bibnamefont {Golez}}, \bibinfo
  {author} {\bibfnamefont {P.}~\bibnamefont {Werner}}, \ and\ \bibinfo {author}
  {\bibfnamefont {M.}~\bibnamefont {Eckstein}},\ }\href {\doibase
  10.1103/PhysRevB.102.165136} {\bibfield  {journal} {\bibinfo  {journal}
  {Phys. Rev. B}\ }\textbf {\bibinfo {volume} {102}},\ \bibinfo {pages}
  {165136} (\bibinfo {year} {2020})}\BibitemShut {NoStop}%
\bibitem [{\citenamefont {Peronaci}\ \emph {et~al.}(2020)\citenamefont
  {Peronaci}, \citenamefont {Parcollet},\ and\ \citenamefont
  {Schir\'o}}]{peronaci2020enhancement}%
  \BibitemOpen
  \bibfield  {author} {\bibinfo {author} {\bibfnamefont {F.}~\bibnamefont
  {Peronaci}}, \bibinfo {author} {\bibfnamefont {O.}~\bibnamefont {Parcollet}},
  \ and\ \bibinfo {author} {\bibfnamefont {M.}~\bibnamefont {Schir\'o}},\
  }\href {\doibase 10.1103/PhysRevB.101.161101} {\bibfield  {journal} {\bibinfo
   {journal} {Phys. Rev. B}\ }\textbf {\bibinfo {volume} {101}},\ \bibinfo
  {pages} {161101} (\bibinfo {year} {2020})}\BibitemShut {NoStop}%
\bibitem [{\citenamefont {Peronaci}\ \emph {et~al.}(2015)\citenamefont
  {Peronaci}, \citenamefont {Schir\'o},\ and\ \citenamefont
  {Capone}}]{peronaci2015transient}%
  \BibitemOpen
  \bibfield  {author} {\bibinfo {author} {\bibfnamefont {F.}~\bibnamefont
  {Peronaci}}, \bibinfo {author} {\bibfnamefont {M.}~\bibnamefont {Schir\'o}},
  \ and\ \bibinfo {author} {\bibfnamefont {M.}~\bibnamefont {Capone}},\ }\href
  {\doibase 10.1103/PhysRevLett.115.257001} {\bibfield  {journal} {\bibinfo
  {journal} {Phys. Rev. Lett.}\ }\textbf {\bibinfo {volume} {115}},\ \bibinfo
  {pages} {257001} (\bibinfo {year} {2015})}\BibitemShut {NoStop}%
\bibitem [{\citenamefont {Poletti}\ \emph {et~al.}(2012)\citenamefont
  {Poletti}, \citenamefont {Bernier}, \citenamefont {Georges},\ and\
  \citenamefont {Kollath}}]{poletti2012interaction}%
  \BibitemOpen
  \bibfield  {author} {\bibinfo {author} {\bibfnamefont {D.}~\bibnamefont
  {Poletti}}, \bibinfo {author} {\bibfnamefont {J.-S.}\ \bibnamefont
  {Bernier}}, \bibinfo {author} {\bibfnamefont {A.}~\bibnamefont {Georges}}, \
  and\ \bibinfo {author} {\bibfnamefont {C.}~\bibnamefont {Kollath}},\ }\href
  {\doibase 10.1103/PhysRevLett.109.045302} {\bibfield  {journal} {\bibinfo
  {journal} {Phys. Rev. Lett.}\ }\textbf {\bibinfo {volume} {109}},\ \bibinfo
  {pages} {045302} (\bibinfo {year} {2012})}\BibitemShut {NoStop}%
\bibitem [{\citenamefont {Poletti}\ \emph {et~al.}(2013)\citenamefont
  {Poletti}, \citenamefont {Barmettler}, \citenamefont {Georges},\ and\
  \citenamefont {Kollath}}]{poletti2013emergence}%
  \BibitemOpen
  \bibfield  {author} {\bibinfo {author} {\bibfnamefont {D.}~\bibnamefont
  {Poletti}}, \bibinfo {author} {\bibfnamefont {P.}~\bibnamefont {Barmettler}},
  \bibinfo {author} {\bibfnamefont {A.}~\bibnamefont {Georges}}, \ and\
  \bibinfo {author} {\bibfnamefont {C.}~\bibnamefont {Kollath}},\ }\href
  {\doibase 10.1103/PhysRevLett.111.195301} {\bibfield  {journal} {\bibinfo
  {journal} {Phys. Rev. Lett.}\ }\textbf {\bibinfo {volume} {111}},\ \bibinfo
  {pages} {195301} (\bibinfo {year} {2013})}\BibitemShut {NoStop}%
\bibitem [{\citenamefont {Sciolla}\ \emph {et~al.}(2015)\citenamefont
  {Sciolla}, \citenamefont {Poletti},\ and\ \citenamefont
  {Kollath}}]{sciolla2015twotime}%
  \BibitemOpen
  \bibfield  {author} {\bibinfo {author} {\bibfnamefont {B.}~\bibnamefont
  {Sciolla}}, \bibinfo {author} {\bibfnamefont {D.}~\bibnamefont {Poletti}}, \
  and\ \bibinfo {author} {\bibfnamefont {C.}~\bibnamefont {Kollath}},\ }\href
  {\doibase 10.1103/PhysRevLett.114.170401} {\bibfield  {journal} {\bibinfo
  {journal} {Phys. Rev. Lett.}\ }\textbf {\bibinfo {volume} {114}},\ \bibinfo
  {pages} {170401} (\bibinfo {year} {2015})}\BibitemShut {NoStop}%
\bibitem [{\citenamefont {Pan}\ \emph {et~al.}(2020)\citenamefont {Pan},
  \citenamefont {Chen}, \citenamefont {Chen},\ and\ \citenamefont
  {Zhai}}]{pan2020nonhermitian}%
  \BibitemOpen
  \bibfield  {author} {\bibinfo {author} {\bibfnamefont {L.}~\bibnamefont
  {Pan}}, \bibinfo {author} {\bibfnamefont {X.}~\bibnamefont {Chen}}, \bibinfo
  {author} {\bibfnamefont {Y.}~\bibnamefont {Chen}}, \ and\ \bibinfo {author}
  {\bibfnamefont {H.}~\bibnamefont {Zhai}},\ }\href {\doibase
  10.1038/s41567-020-0889-6} {\bibfield  {journal} {\bibinfo  {journal} {Nature
  Physics}\ }\textbf {\bibinfo {volume} {16}},\ \bibinfo {pages} {767}
  (\bibinfo {year} {2020})}\BibitemShut {NoStop}%
\bibitem [{\citenamefont {Gerbier}\ and\ \citenamefont
  {Castin}(2010)}]{gerbier2010heating}%
  \BibitemOpen
  \bibfield  {author} {\bibinfo {author} {\bibfnamefont {F.}~\bibnamefont
  {Gerbier}}\ and\ \bibinfo {author} {\bibfnamefont {Y.}~\bibnamefont
  {Castin}},\ }\href {\doibase 10.1103/PhysRevA.82.013615} {\bibfield
  {journal} {\bibinfo  {journal} {Phys. Rev. A}\ }\textbf {\bibinfo {volume}
  {82}},\ \bibinfo {pages} {013615} (\bibinfo {year} {2010})}\BibitemShut
  {NoStop}%
\bibitem [{\citenamefont {Bouganne}\ \emph {et~al.}(2020)\citenamefont
  {Bouganne}, \citenamefont {Bosch~Aguilera}, \citenamefont {Ghermaoui},
  \citenamefont {Beugnon},\ and\ \citenamefont {Gerbier}}]{bouganne2020}%
  \BibitemOpen
  \bibfield  {author} {\bibinfo {author} {\bibfnamefont {R.}~\bibnamefont
  {Bouganne}}, \bibinfo {author} {\bibfnamefont {M.}~\bibnamefont
  {Bosch~Aguilera}}, \bibinfo {author} {\bibfnamefont {A.}~\bibnamefont
  {Ghermaoui}}, \bibinfo {author} {\bibfnamefont {J.}~\bibnamefont {Beugnon}},
  \ and\ \bibinfo {author} {\bibfnamefont {F.}~\bibnamefont {Gerbier}},\ }\href
  {\doibase 10.1038/s41567-019-0678-2} {\bibfield  {journal} {\bibinfo
  {journal} {Nature Physics}\ }\textbf {\bibinfo {volume} {16}},\ \bibinfo
  {pages} {21} (\bibinfo {year} {2020})}\BibitemShut {NoStop}%
\bibitem [{\citenamefont {Tindall}\ \emph {et~al.}(2019)\citenamefont
  {Tindall}, \citenamefont {Bu\ifmmode~\check{c}\else \v{c}\fi{}a},
  \citenamefont {Coulthard},\ and\ \citenamefont
  {Jaksch}}]{tindall2019heating}%
  \BibitemOpen
  \bibfield  {author} {\bibinfo {author} {\bibfnamefont {J.}~\bibnamefont
  {Tindall}}, \bibinfo {author} {\bibfnamefont {B.}~\bibnamefont
  {Bu\ifmmode~\check{c}\else \v{c}\fi{}a}}, \bibinfo {author} {\bibfnamefont
  {J.~R.}\ \bibnamefont {Coulthard}}, \ and\ \bibinfo {author} {\bibfnamefont
  {D.}~\bibnamefont {Jaksch}},\ }\href {\doibase
  10.1103/PhysRevLett.123.030603} {\bibfield  {journal} {\bibinfo  {journal}
  {Phys. Rev. Lett.}\ }\textbf {\bibinfo {volume} {123}},\ \bibinfo {pages}
  {030603} (\bibinfo {year} {2019})}\BibitemShut {NoStop}%
\bibitem [{\citenamefont {Nakagawa}\ \emph {et~al.}(2021)\citenamefont
  {Nakagawa}, \citenamefont {Tsuji}, \citenamefont {Kawakami},\ and\
  \citenamefont {Ueda}}]{nakagawa2021etapairing}%
  \BibitemOpen
  \bibfield  {author} {\bibinfo {author} {\bibfnamefont {M.}~\bibnamefont
  {Nakagawa}}, \bibinfo {author} {\bibfnamefont {N.}~\bibnamefont {Tsuji}},
  \bibinfo {author} {\bibfnamefont {N.}~\bibnamefont {Kawakami}}, \ and\
  \bibinfo {author} {\bibfnamefont {M.}~\bibnamefont {Ueda}},\ }\href {\doibase
  10.48550/ARXIV.2103.13624} {\enquote {\bibinfo {title} {$\eta$ pairing of
  light-emitting fermions: Nonequilibrium pairing mechanism at high
  temperatures},}\ } (\bibinfo {year} {2021})\BibitemShut {NoStop}%
\bibitem [{\citenamefont {Kantian}\ \emph {et~al.}(2009)\citenamefont
  {Kantian}, \citenamefont {Dalmonte}, \citenamefont {Diehl}, \citenamefont
  {Hofstetter}, \citenamefont {Zoller},\ and\ \citenamefont
  {Daley}}]{kantian2009atomic}%
  \BibitemOpen
  \bibfield  {author} {\bibinfo {author} {\bibfnamefont {A.}~\bibnamefont
  {Kantian}}, \bibinfo {author} {\bibfnamefont {M.}~\bibnamefont {Dalmonte}},
  \bibinfo {author} {\bibfnamefont {S.}~\bibnamefont {Diehl}}, \bibinfo
  {author} {\bibfnamefont {W.}~\bibnamefont {Hofstetter}}, \bibinfo {author}
  {\bibfnamefont {P.}~\bibnamefont {Zoller}}, \ and\ \bibinfo {author}
  {\bibfnamefont {A.~J.}\ \bibnamefont {Daley}},\ }\href {\doibase
  10.1103/PhysRevLett.103.240401} {\bibfield  {journal} {\bibinfo  {journal}
  {Phys. Rev. Lett.}\ }\textbf {\bibinfo {volume} {103}},\ \bibinfo {pages}
  {240401} (\bibinfo {year} {2009})}\BibitemShut {NoStop}%
\bibitem [{\citenamefont {Syassen}\ \emph {et~al.}(2008)\citenamefont
  {Syassen}, \citenamefont {Bauer}, \citenamefont {Lettner}, \citenamefont
  {Volz}, \citenamefont {Dietze}, \citenamefont {Garc{\'\i}a-Ripoll},
  \citenamefont {Cirac}, \citenamefont {Rempe},\ and\ \citenamefont
  {D{\"u}rr}}]{Syassen1329}%
  \BibitemOpen
  \bibfield  {author} {\bibinfo {author} {\bibfnamefont {N.}~\bibnamefont
  {Syassen}}, \bibinfo {author} {\bibfnamefont {D.~M.}\ \bibnamefont {Bauer}},
  \bibinfo {author} {\bibfnamefont {M.}~\bibnamefont {Lettner}}, \bibinfo
  {author} {\bibfnamefont {T.}~\bibnamefont {Volz}}, \bibinfo {author}
  {\bibfnamefont {D.}~\bibnamefont {Dietze}}, \bibinfo {author} {\bibfnamefont
  {J.~J.}\ \bibnamefont {Garc{\'\i}a-Ripoll}}, \bibinfo {author} {\bibfnamefont
  {J.~I.}\ \bibnamefont {Cirac}}, \bibinfo {author} {\bibfnamefont
  {G.}~\bibnamefont {Rempe}}, \ and\ \bibinfo {author} {\bibfnamefont
  {S.}~\bibnamefont {D{\"u}rr}},\ }\href {\doibase 10.1126/science.1155309}
  {\bibfield  {journal} {\bibinfo  {journal} {Science}\ }\textbf {\bibinfo
  {volume} {320}},\ \bibinfo {pages} {1329} (\bibinfo {year} {2008})},\ \Eprint
  {http://arxiv.org/abs/https://science.sciencemag.org/content/320/5881/1329.full.pdf}
  {https://science.sciencemag.org/content/320/5881/1329.full.pdf} \BibitemShut
  {NoStop}%
\bibitem [{\citenamefont {Tomita}\ \emph {et~al.}(2017)\citenamefont {Tomita},
  \citenamefont {Nakajima}, \citenamefont {Danshita}, \citenamefont {Takasu},\
  and\ \citenamefont {Takahashi}}]{TomitaEtAlScienceAdv17}%
  \BibitemOpen
  \bibfield  {author} {\bibinfo {author} {\bibfnamefont {T.}~\bibnamefont
  {Tomita}}, \bibinfo {author} {\bibfnamefont {S.}~\bibnamefont {Nakajima}},
  \bibinfo {author} {\bibfnamefont {I.}~\bibnamefont {Danshita}}, \bibinfo
  {author} {\bibfnamefont {Y.}~\bibnamefont {Takasu}}, \ and\ \bibinfo {author}
  {\bibfnamefont {Y.}~\bibnamefont {Takahashi}},\ }\href {\doibase
  10.1126/sciadv.1701513} {\bibfield  {journal} {\bibinfo  {journal} {Science
  Advances}\ }\textbf {\bibinfo {volume} {3}} (\bibinfo {year} {2017}),\
  10.1126/sciadv.1701513}\BibitemShut {NoStop}%
\bibitem [{\citenamefont {Sponselee}\ \emph {et~al.}(2018)\citenamefont
  {Sponselee}, \citenamefont {Freystatzky}, \citenamefont {Abeln},
  \citenamefont {Diem}, \citenamefont {Hundt}, \citenamefont {Kochanke},
  \citenamefont {Ponath}, \citenamefont {Santra}, \citenamefont {Mathey},
  \citenamefont {Sengstock},\ and\ \citenamefont {Becker}}]{Sponselee_2018}%
  \BibitemOpen
  \bibfield  {author} {\bibinfo {author} {\bibfnamefont {K.}~\bibnamefont
  {Sponselee}}, \bibinfo {author} {\bibfnamefont {L.}~\bibnamefont
  {Freystatzky}}, \bibinfo {author} {\bibfnamefont {B.}~\bibnamefont {Abeln}},
  \bibinfo {author} {\bibfnamefont {M.}~\bibnamefont {Diem}}, \bibinfo {author}
  {\bibfnamefont {B.}~\bibnamefont {Hundt}}, \bibinfo {author} {\bibfnamefont
  {A.}~\bibnamefont {Kochanke}}, \bibinfo {author} {\bibfnamefont
  {T.}~\bibnamefont {Ponath}}, \bibinfo {author} {\bibfnamefont
  {B.}~\bibnamefont {Santra}}, \bibinfo {author} {\bibfnamefont
  {L.}~\bibnamefont {Mathey}}, \bibinfo {author} {\bibfnamefont
  {K.}~\bibnamefont {Sengstock}}, \ and\ \bibinfo {author} {\bibfnamefont
  {C.}~\bibnamefont {Becker}},\ }\href {\doibase 10.1088/2058-9565/aadccd}
  {\bibfield  {journal} {\bibinfo  {journal} {Quantum Science and Technology}\
  }\textbf {\bibinfo {volume} {4}},\ \bibinfo {pages} {014002} (\bibinfo {year}
  {2018})}\BibitemShut {NoStop}%
\bibitem [{\citenamefont {Honda}\ \emph {et~al.}(2022)\citenamefont {Honda},
  \citenamefont {Taie}, \citenamefont {Takasu}, \citenamefont {Nishizawa},
  \citenamefont {Nakagawa},\ and\ \citenamefont
  {Takahashi}}]{honda2022observation}%
  \BibitemOpen
  \bibfield  {author} {\bibinfo {author} {\bibfnamefont {K.}~\bibnamefont
  {Honda}}, \bibinfo {author} {\bibfnamefont {S.}~\bibnamefont {Taie}},
  \bibinfo {author} {\bibfnamefont {Y.}~\bibnamefont {Takasu}}, \bibinfo
  {author} {\bibfnamefont {N.}~\bibnamefont {Nishizawa}}, \bibinfo {author}
  {\bibfnamefont {M.}~\bibnamefont {Nakagawa}}, \ and\ \bibinfo {author}
  {\bibfnamefont {Y.}~\bibnamefont {Takahashi}},\ }\href {\doibase
  10.48550/ARXIV.2205.13162} {\enquote {\bibinfo {title} {Observation of the
  sign reversal of the magnetic correlation in a driven-dissipative
  fermi-hubbard system},}\ } (\bibinfo {year} {2022})\BibitemShut {NoStop}%
\bibitem [{\citenamefont {Wang}\ \emph {et~al.}(2022)\citenamefont {Wang},
  \citenamefont {Liu},\ and\ \citenamefont {Shi}}]{wang2022complex}%
  \BibitemOpen
  \bibfield  {author} {\bibinfo {author} {\bibfnamefont {C.}~\bibnamefont
  {Wang}}, \bibinfo {author} {\bibfnamefont {C.}~\bibnamefont {Liu}}, \ and\
  \bibinfo {author} {\bibfnamefont {Z.-Y.}\ \bibnamefont {Shi}},\ }\href
  {\doibase 10.1103/PhysRevLett.129.203401} {\bibfield  {journal} {\bibinfo
  {journal} {Phys. Rev. Lett.}\ }\textbf {\bibinfo {volume} {129}},\ \bibinfo
  {pages} {203401} (\bibinfo {year} {2022})}\BibitemShut {NoStop}%
\bibitem [{\citenamefont {Zhou}\ and\ \citenamefont
  {Cui}(2021)}]{zhou2021effective}%
  \BibitemOpen
  \bibfield  {author} {\bibinfo {author} {\bibfnamefont {L.}~\bibnamefont
  {Zhou}}\ and\ \bibinfo {author} {\bibfnamefont {X.}~\bibnamefont {Cui}},\
  }\href {\doibase 10.1103/PhysRevResearch.3.043225} {\bibfield  {journal}
  {\bibinfo  {journal} {Phys. Rev. Res.}\ }\textbf {\bibinfo {volume} {3}},\
  \bibinfo {pages} {043225} (\bibinfo {year} {2021})}\BibitemShut {NoStop}%
\bibitem [{\citenamefont {Sandner}\ \emph {et~al.}(2011)\citenamefont
  {Sandner}, \citenamefont {M\"uller}, \citenamefont {Daley},\ and\
  \citenamefont {Zoller}}]{sandner2011spatial}%
  \BibitemOpen
  \bibfield  {author} {\bibinfo {author} {\bibfnamefont {R.~M.}\ \bibnamefont
  {Sandner}}, \bibinfo {author} {\bibfnamefont {M.}~\bibnamefont {M\"uller}},
  \bibinfo {author} {\bibfnamefont {A.~J.}\ \bibnamefont {Daley}}, \ and\
  \bibinfo {author} {\bibfnamefont {P.}~\bibnamefont {Zoller}},\ }\href
  {\doibase 10.1103/PhysRevA.84.043825} {\bibfield  {journal} {\bibinfo
  {journal} {Phys. Rev. A}\ }\textbf {\bibinfo {volume} {84}},\ \bibinfo
  {pages} {043825} (\bibinfo {year} {2011})}\BibitemShut {NoStop}%
\bibitem [{\citenamefont {Zhang}\ \emph {et~al.}(2015)\citenamefont {Zhang},
  \citenamefont {Cheng}, \citenamefont {Zhai},\ and\ \citenamefont
  {Zhang}}]{zhang2015orbital}%
  \BibitemOpen
  \bibfield  {author} {\bibinfo {author} {\bibfnamefont {R.}~\bibnamefont
  {Zhang}}, \bibinfo {author} {\bibfnamefont {Y.}~\bibnamefont {Cheng}},
  \bibinfo {author} {\bibfnamefont {H.}~\bibnamefont {Zhai}}, \ and\ \bibinfo
  {author} {\bibfnamefont {P.}~\bibnamefont {Zhang}},\ }\href {\doibase
  10.1103/PhysRevLett.115.135301} {\bibfield  {journal} {\bibinfo  {journal}
  {Phys. Rev. Lett.}\ }\textbf {\bibinfo {volume} {115}},\ \bibinfo {pages}
  {135301} (\bibinfo {year} {2015})}\BibitemShut {NoStop}%
\bibitem [{\citenamefont {Pagano}\ \emph {et~al.}(2015)\citenamefont {Pagano},
  \citenamefont {Mancini}, \citenamefont {Cappellini}, \citenamefont {Livi},
  \citenamefont {Sias}, \citenamefont {Catani}, \citenamefont {Inguscio},\ and\
  \citenamefont {Fallani}}]{pagano2015strongly}%
  \BibitemOpen
  \bibfield  {author} {\bibinfo {author} {\bibfnamefont {G.}~\bibnamefont
  {Pagano}}, \bibinfo {author} {\bibfnamefont {M.}~\bibnamefont {Mancini}},
  \bibinfo {author} {\bibfnamefont {G.}~\bibnamefont {Cappellini}}, \bibinfo
  {author} {\bibfnamefont {L.}~\bibnamefont {Livi}}, \bibinfo {author}
  {\bibfnamefont {C.}~\bibnamefont {Sias}}, \bibinfo {author} {\bibfnamefont
  {J.}~\bibnamefont {Catani}}, \bibinfo {author} {\bibfnamefont
  {M.}~\bibnamefont {Inguscio}}, \ and\ \bibinfo {author} {\bibfnamefont
  {L.}~\bibnamefont {Fallani}},\ }\href {\doibase
  10.1103/PhysRevLett.115.265301} {\bibfield  {journal} {\bibinfo  {journal}
  {Phys. Rev. Lett.}\ }\textbf {\bibinfo {volume} {115}},\ \bibinfo {pages}
  {265301} (\bibinfo {year} {2015})}\BibitemShut {NoStop}%
\bibitem [{\citenamefont {Foss-Feig}\ \emph {et~al.}(2012)\citenamefont
  {Foss-Feig}, \citenamefont {Daley}, \citenamefont {Thompson},\ and\
  \citenamefont {Rey}}]{fossfeig2012steady}%
  \BibitemOpen
  \bibfield  {author} {\bibinfo {author} {\bibfnamefont {M.}~\bibnamefont
  {Foss-Feig}}, \bibinfo {author} {\bibfnamefont {A.~J.}\ \bibnamefont
  {Daley}}, \bibinfo {author} {\bibfnamefont {J.~K.}\ \bibnamefont {Thompson}},
  \ and\ \bibinfo {author} {\bibfnamefont {A.~M.}\ \bibnamefont {Rey}},\ }\href
  {\doibase 10.1103/PhysRevLett.109.230501} {\bibfield  {journal} {\bibinfo
  {journal} {Phys. Rev. Lett.}\ }\textbf {\bibinfo {volume} {109}},\ \bibinfo
  {pages} {230501} (\bibinfo {year} {2012})}\BibitemShut {NoStop}%
\bibitem [{\citenamefont {Rosso}\ \emph {et~al.}(2021)\citenamefont {Rosso},
  \citenamefont {Rossini}, \citenamefont {Biella},\ and\ \citenamefont
  {Mazza}}]{rosso2021onedimensional}%
  \BibitemOpen
  \bibfield  {author} {\bibinfo {author} {\bibfnamefont {L.}~\bibnamefont
  {Rosso}}, \bibinfo {author} {\bibfnamefont {D.}~\bibnamefont {Rossini}},
  \bibinfo {author} {\bibfnamefont {A.}~\bibnamefont {Biella}}, \ and\ \bibinfo
  {author} {\bibfnamefont {L.}~\bibnamefont {Mazza}},\ }\href {\doibase
  10.1103/PhysRevA.104.053305} {\bibfield  {journal} {\bibinfo  {journal}
  {Phys. Rev. A}\ }\textbf {\bibinfo {volume} {104}},\ \bibinfo {pages}
  {053305} (\bibinfo {year} {2021})}\BibitemShut {NoStop}%
\bibitem [{\citenamefont {Misra}\ and\ \citenamefont
  {Sudarshan}(1977)}]{misra1977a}%
  \BibitemOpen
  \bibfield  {author} {\bibinfo {author} {\bibfnamefont {B.}~\bibnamefont
  {Misra}}\ and\ \bibinfo {author} {\bibfnamefont {E.~C.~G.}\ \bibnamefont
  {Sudarshan}},\ }\href@noop {} {\bibfield  {journal} {\bibinfo  {journal}
  {Journal of Mathematical Physics}\ }\textbf {\bibinfo {volume} {18}},\
  \bibinfo {pages} {756} (\bibinfo {year} {1977})}\BibitemShut {NoStop}%
\bibitem [{\citenamefont {{Garc{\'i}a-Ripoll}}\ \emph
  {et~al.}(2009)\citenamefont {{Garc{\'i}a-Ripoll}}, \citenamefont {D{\"u}rr},
  \citenamefont {Syassen}, \citenamefont {Bauer}, \citenamefont {Lettner},
  \citenamefont {Rempe},\ and\ \citenamefont {Cirac}}]{garcia-ripoll2009}%
  \BibitemOpen
  \bibfield  {author} {\bibinfo {author} {\bibfnamefont {J.~J.}\ \bibnamefont
  {{Garc{\'i}a-Ripoll}}}, \bibinfo {author} {\bibfnamefont {S.}~\bibnamefont
  {D{\"u}rr}}, \bibinfo {author} {\bibfnamefont {N.}~\bibnamefont {Syassen}},
  \bibinfo {author} {\bibfnamefont {D.~M.}\ \bibnamefont {Bauer}}, \bibinfo
  {author} {\bibfnamefont {M.}~\bibnamefont {Lettner}}, \bibinfo {author}
  {\bibfnamefont {G.}~\bibnamefont {Rempe}}, \ and\ \bibinfo {author}
  {\bibfnamefont {J.~I.}\ \bibnamefont {Cirac}},\ }\href@noop {} {\bibfield
  {journal} {\bibinfo  {journal} {New Journal of Physics}\ }\textbf {\bibinfo
  {volume} {11}},\ \bibinfo {pages} {013053} (\bibinfo {year}
  {2009})}\BibitemShut {NoStop}%
\bibitem [{\citenamefont {Zhu}\ \emph {et~al.}(2014)\citenamefont {Zhu},
  \citenamefont {Gadway}, \citenamefont {Foss-Feig}, \citenamefont
  {Schachenmayer}, \citenamefont {Wall}, \citenamefont {Hazzard}, \citenamefont
  {Yan}, \citenamefont {Moses}, \citenamefont {Covey}, \citenamefont {Jin},
  \citenamefont {Ye}, \citenamefont {Holland},\ and\ \citenamefont
  {Rey}}]{zhu2014suppressing}%
  \BibitemOpen
  \bibfield  {author} {\bibinfo {author} {\bibfnamefont {B.}~\bibnamefont
  {Zhu}}, \bibinfo {author} {\bibfnamefont {B.}~\bibnamefont {Gadway}},
  \bibinfo {author} {\bibfnamefont {M.}~\bibnamefont {Foss-Feig}}, \bibinfo
  {author} {\bibfnamefont {J.}~\bibnamefont {Schachenmayer}}, \bibinfo {author}
  {\bibfnamefont {M.~L.}\ \bibnamefont {Wall}}, \bibinfo {author}
  {\bibfnamefont {K.~R.~A.}\ \bibnamefont {Hazzard}}, \bibinfo {author}
  {\bibfnamefont {B.}~\bibnamefont {Yan}}, \bibinfo {author} {\bibfnamefont
  {S.~A.}\ \bibnamefont {Moses}}, \bibinfo {author} {\bibfnamefont {J.~P.}\
  \bibnamefont {Covey}}, \bibinfo {author} {\bibfnamefont {D.~S.}\ \bibnamefont
  {Jin}}, \bibinfo {author} {\bibfnamefont {J.}~\bibnamefont {Ye}}, \bibinfo
  {author} {\bibfnamefont {M.}~\bibnamefont {Holland}}, \ and\ \bibinfo
  {author} {\bibfnamefont {A.~M.}\ \bibnamefont {Rey}},\ }\href {\doibase
  10.1103/PhysRevLett.112.070404} {\bibfield  {journal} {\bibinfo  {journal}
  {Phys. Rev. Lett.}\ }\textbf {\bibinfo {volume} {112}},\ \bibinfo {pages}
  {070404} (\bibinfo {year} {2014})}\BibitemShut {NoStop}%
\bibitem [{\citenamefont {Fr{\"o}ml}\ \emph {et~al.}(2019)\citenamefont
  {Fr{\"o}ml}, \citenamefont {Chiocchetta}, \citenamefont {Kollath},\ and\
  \citenamefont {Diehl}}]{Froml2019}%
  \BibitemOpen
  \bibfield  {author} {\bibinfo {author} {\bibfnamefont {H.}~\bibnamefont
  {Fr{\"o}ml}}, \bibinfo {author} {\bibfnamefont {A.}~\bibnamefont
  {Chiocchetta}}, \bibinfo {author} {\bibfnamefont {C.}~\bibnamefont
  {Kollath}}, \ and\ \bibinfo {author} {\bibfnamefont {S.}~\bibnamefont
  {Diehl}},\ }\href {\doibase 10.1103/PhysRevLett.122.040402} {\bibfield
  {journal} {\bibinfo  {journal} {Physical Review Letters}\ }\textbf {\bibinfo
  {volume} {122}},\ \bibinfo {pages} {040402} (\bibinfo {year}
  {2019})}\BibitemShut {NoStop}%
\bibitem [{\citenamefont {Nakagawa}\ \emph {et~al.}(2020)\citenamefont
  {Nakagawa}, \citenamefont {Tsuji}, \citenamefont {Kawakami},\ and\
  \citenamefont {Ueda}}]{NakagawaEtAlPRL20}%
  \BibitemOpen
  \bibfield  {author} {\bibinfo {author} {\bibfnamefont {M.}~\bibnamefont
  {Nakagawa}}, \bibinfo {author} {\bibfnamefont {N.}~\bibnamefont {Tsuji}},
  \bibinfo {author} {\bibfnamefont {N.}~\bibnamefont {Kawakami}}, \ and\
  \bibinfo {author} {\bibfnamefont {M.}~\bibnamefont {Ueda}},\ }\href {\doibase
  10.1103/PhysRevLett.124.147203} {\bibfield  {journal} {\bibinfo  {journal}
  {Phys. Rev. Lett.}\ }\textbf {\bibinfo {volume} {124}},\ \bibinfo {pages}
  {147203} (\bibinfo {year} {2020})}\BibitemShut {NoStop}%
\bibitem [{\citenamefont {Rossini}\ \emph {et~al.}(2021)\citenamefont
  {Rossini}, \citenamefont {Ghermaoui}, \citenamefont {Aguilera}, \citenamefont
  {Vatr\'e}, \citenamefont {Bouganne}, \citenamefont {Beugnon}, \citenamefont
  {Gerbier},\ and\ \citenamefont {Mazza}}]{rossini2021strong}%
  \BibitemOpen
  \bibfield  {author} {\bibinfo {author} {\bibfnamefont {D.}~\bibnamefont
  {Rossini}}, \bibinfo {author} {\bibfnamefont {A.}~\bibnamefont {Ghermaoui}},
  \bibinfo {author} {\bibfnamefont {M.~B.}\ \bibnamefont {Aguilera}}, \bibinfo
  {author} {\bibfnamefont {R.}~\bibnamefont {Vatr\'e}}, \bibinfo {author}
  {\bibfnamefont {R.}~\bibnamefont {Bouganne}}, \bibinfo {author}
  {\bibfnamefont {J.}~\bibnamefont {Beugnon}}, \bibinfo {author} {\bibfnamefont
  {F.}~\bibnamefont {Gerbier}}, \ and\ \bibinfo {author} {\bibfnamefont
  {L.}~\bibnamefont {Mazza}},\ }\href {\doibase 10.1103/PhysRevA.103.L060201}
  {\bibfield  {journal} {\bibinfo  {journal} {Phys. Rev. A}\ }\textbf {\bibinfo
  {volume} {103}},\ \bibinfo {pages} {L060201} (\bibinfo {year}
  {2021})}\BibitemShut {NoStop}%
\bibitem [{\citenamefont {Scarlatella}\ \emph {et~al.}(2021)\citenamefont
  {Scarlatella}, \citenamefont {Clerk}, \citenamefont {Fazio},\ and\
  \citenamefont {Schir\'o}}]{scarlatella2021dynamical}%
  \BibitemOpen
  \bibfield  {author} {\bibinfo {author} {\bibfnamefont {O.}~\bibnamefont
  {Scarlatella}}, \bibinfo {author} {\bibfnamefont {A.~A.}\ \bibnamefont
  {Clerk}}, \bibinfo {author} {\bibfnamefont {R.}~\bibnamefont {Fazio}}, \ and\
  \bibinfo {author} {\bibfnamefont {M.}~\bibnamefont {Schir\'o}},\ }\href
  {\doibase 10.1103/PhysRevX.11.031018} {\bibfield  {journal} {\bibinfo
  {journal} {Phys. Rev. X}\ }\textbf {\bibinfo {volume} {11}},\ \bibinfo
  {pages} {031018} (\bibinfo {year} {2021})}\BibitemShut {NoStop}%
\bibitem [{\citenamefont {Biella}\ and\ \citenamefont
  {Schir{\'{o}}}(2021)}]{Biella2021manybodyquantumzeno}%
  \BibitemOpen
  \bibfield  {author} {\bibinfo {author} {\bibfnamefont {A.}~\bibnamefont
  {Biella}}\ and\ \bibinfo {author} {\bibfnamefont {M.}~\bibnamefont
  {Schir{\'{o}}}},\ }\href {\doibase 10.22331/q-2021-08-19-528} {\bibfield
  {journal} {\bibinfo  {journal} {{Quantum}}\ }\textbf {\bibinfo {volume}
  {5}},\ \bibinfo {pages} {528} (\bibinfo {year} {2021})}\BibitemShut {NoStop}%
\bibitem [{\citenamefont {Rosso}\ \emph
  {et~al.}(2022{\natexlab{a}})\citenamefont {Rosso}, \citenamefont {Biella},\
  and\ \citenamefont {Mazza}}]{rosso2022the}%
  \BibitemOpen
  \bibfield  {author} {\bibinfo {author} {\bibfnamefont {L.}~\bibnamefont
  {Rosso}}, \bibinfo {author} {\bibfnamefont {A.}~\bibnamefont {Biella}}, \
  and\ \bibinfo {author} {\bibfnamefont {L.}~\bibnamefont {Mazza}},\ }\href
  {\doibase 10.21468/SciPostPhys.12.1.044} {\bibfield  {journal} {\bibinfo
  {journal} {SciPost Phys.}\ }\textbf {\bibinfo {volume} {12}},\ \bibinfo
  {pages} {44} (\bibinfo {year} {2022}{\natexlab{a}})}\BibitemShut {NoStop}%
\bibitem [{\citenamefont {Rosso}\ \emph
  {et~al.}(2022{\natexlab{b}})\citenamefont {Rosso}, \citenamefont {Biella},
  \citenamefont {De~Nardis},\ and\ \citenamefont
  {Mazza}}]{rosso2022adynamical}%
  \BibitemOpen
  \bibfield  {author} {\bibinfo {author} {\bibfnamefont {L.}~\bibnamefont
  {Rosso}}, \bibinfo {author} {\bibfnamefont {A.}~\bibnamefont {Biella}},
  \bibinfo {author} {\bibfnamefont {J.}~\bibnamefont {De~Nardis}}, \ and\
  \bibinfo {author} {\bibfnamefont {L.}~\bibnamefont {Mazza}},\ }\href
  {\doibase 10.48550/ARXIV.2206.06837} {\enquote {\bibinfo {title} {A dynamical
  theory for one-dimensional fermions with strong two-body losses: universal
  non-hermitian zeno physics and spin-charge separation},}\ } (\bibinfo {year}
  {2022}{\natexlab{b}})\BibitemShut {NoStop}%
\bibitem [{\citenamefont {Secl\`{\i}}\ \emph {et~al.}(2022)\citenamefont
  {Secl\`{\i}}, \citenamefont {Capone},\ and\ \citenamefont
  {Schir\`o}}]{secli2022steady}%
  \BibitemOpen
  \bibfield  {author} {\bibinfo {author} {\bibfnamefont {M.}~\bibnamefont
  {Secl\`{\i}}}, \bibinfo {author} {\bibfnamefont {M.}~\bibnamefont {Capone}},
  \ and\ \bibinfo {author} {\bibfnamefont {M.}~\bibnamefont {Schir\`o}},\
  }\href {\doibase 10.1103/PhysRevA.106.013707} {\bibfield  {journal} {\bibinfo
   {journal} {Phys. Rev. A}\ }\textbf {\bibinfo {volume} {106}},\ \bibinfo
  {pages} {013707} (\bibinfo {year} {2022})}\BibitemShut {NoStop}%
\bibitem [{\citenamefont {Mitra}\ \emph {et~al.}(2018)\citenamefont {Mitra},
  \citenamefont {Brown}, \citenamefont {Guardado-Sanchez}, \citenamefont
  {Kondov}, \citenamefont {Devakul}, \citenamefont {Huse}, \citenamefont
  {Schau{\ss}},\ and\ \citenamefont {Bakr}}]{mitra2018quantum}%
  \BibitemOpen
  \bibfield  {author} {\bibinfo {author} {\bibfnamefont {D.}~\bibnamefont
  {Mitra}}, \bibinfo {author} {\bibfnamefont {P.~T.}\ \bibnamefont {Brown}},
  \bibinfo {author} {\bibfnamefont {E.}~\bibnamefont {Guardado-Sanchez}},
  \bibinfo {author} {\bibfnamefont {S.~S.}\ \bibnamefont {Kondov}}, \bibinfo
  {author} {\bibfnamefont {T.}~\bibnamefont {Devakul}}, \bibinfo {author}
  {\bibfnamefont {D.~A.}\ \bibnamefont {Huse}}, \bibinfo {author}
  {\bibfnamefont {P.}~\bibnamefont {Schau{\ss}}}, \ and\ \bibinfo {author}
  {\bibfnamefont {W.~S.}\ \bibnamefont {Bakr}},\ }\href {\doibase
  10.1038/nphys4297} {\bibfield  {journal} {\bibinfo  {journal} {Nature
  Physics}\ }\textbf {\bibinfo {volume} {14}},\ \bibinfo {pages} {173}
  (\bibinfo {year} {2018})}\BibitemShut {NoStop}%
\bibitem [{\citenamefont {Yamamoto}\ \emph {et~al.}(2019)\citenamefont
  {Yamamoto}, \citenamefont {Nakagawa}, \citenamefont {Adachi}, \citenamefont
  {Takasan}, \citenamefont {Ueda},\ and\ \citenamefont
  {Kawakami}}]{YamamotoEtAlPRL19}%
  \BibitemOpen
  \bibfield  {author} {\bibinfo {author} {\bibfnamefont {K.}~\bibnamefont
  {Yamamoto}}, \bibinfo {author} {\bibfnamefont {M.}~\bibnamefont {Nakagawa}},
  \bibinfo {author} {\bibfnamefont {K.}~\bibnamefont {Adachi}}, \bibinfo
  {author} {\bibfnamefont {K.}~\bibnamefont {Takasan}}, \bibinfo {author}
  {\bibfnamefont {M.}~\bibnamefont {Ueda}}, \ and\ \bibinfo {author}
  {\bibfnamefont {N.}~\bibnamefont {Kawakami}},\ }\href {\doibase
  10.1103/PhysRevLett.123.123601} {\bibfield  {journal} {\bibinfo  {journal}
  {Phys. Rev. Lett.}\ }\textbf {\bibinfo {volume} {123}},\ \bibinfo {pages}
  {123601} (\bibinfo {year} {2019})}\BibitemShut {NoStop}%
\bibitem [{\citenamefont {Iskin}(2021)}]{iskin2021nonhermitian}%
  \BibitemOpen
  \bibfield  {author} {\bibinfo {author} {\bibfnamefont {M.}~\bibnamefont
  {Iskin}},\ }\href {\doibase 10.1103/PhysRevA.103.013724} {\bibfield
  {journal} {\bibinfo  {journal} {Phys. Rev. A}\ }\textbf {\bibinfo {volume}
  {103}},\ \bibinfo {pages} {013724} (\bibinfo {year} {2021})}\BibitemShut
  {NoStop}%
\bibitem [{\citenamefont {Yamamoto}\ \emph {et~al.}(2021)\citenamefont
  {Yamamoto}, \citenamefont {Nakagawa}, \citenamefont {Tsuji}, \citenamefont
  {Ueda},\ and\ \citenamefont {Kawakami}}]{yamamoto2021collective}%
  \BibitemOpen
  \bibfield  {author} {\bibinfo {author} {\bibfnamefont {K.}~\bibnamefont
  {Yamamoto}}, \bibinfo {author} {\bibfnamefont {M.}~\bibnamefont {Nakagawa}},
  \bibinfo {author} {\bibfnamefont {N.}~\bibnamefont {Tsuji}}, \bibinfo
  {author} {\bibfnamefont {M.}~\bibnamefont {Ueda}}, \ and\ \bibinfo {author}
  {\bibfnamefont {N.}~\bibnamefont {Kawakami}},\ }\href {\doibase
  10.1103/PhysRevLett.127.055301} {\bibfield  {journal} {\bibinfo  {journal}
  {Phys. Rev. Lett.}\ }\textbf {\bibinfo {volume} {127}},\ \bibinfo {pages}
  {055301} (\bibinfo {year} {2021})}\BibitemShut {NoStop}%
\bibitem [{\citenamefont {Kwon}\ \emph {et~al.}(2020)\citenamefont {Kwon},
  \citenamefont {Pace}, \citenamefont {Panza}, \citenamefont {Inguscio},
  \citenamefont {Zwerger}, \citenamefont {Zaccanti}, \citenamefont {Scazza},\
  and\ \citenamefont {Roati}}]{kwon2020strongly}%
  \BibitemOpen
  \bibfield  {author} {\bibinfo {author} {\bibfnamefont {W.~J.}\ \bibnamefont
  {Kwon}}, \bibinfo {author} {\bibfnamefont {G.~D.}\ \bibnamefont {Pace}},
  \bibinfo {author} {\bibfnamefont {R.}~\bibnamefont {Panza}}, \bibinfo
  {author} {\bibfnamefont {M.}~\bibnamefont {Inguscio}}, \bibinfo {author}
  {\bibfnamefont {W.}~\bibnamefont {Zwerger}}, \bibinfo {author} {\bibfnamefont
  {M.}~\bibnamefont {Zaccanti}}, \bibinfo {author} {\bibfnamefont
  {F.}~\bibnamefont {Scazza}}, \ and\ \bibinfo {author} {\bibfnamefont
  {G.}~\bibnamefont {Roati}},\ }\href {\doibase 10.1126/science.aaz2463}
  {\bibfield  {journal} {\bibinfo  {journal} {Science}\ }\textbf {\bibinfo
  {volume} {369}},\ \bibinfo {pages} {84} (\bibinfo {year} {2020})},\ \Eprint
  {http://arxiv.org/abs/https://www.science.org/doi/pdf/10.1126/science.aaz2463}
  {https://www.science.org/doi/pdf/10.1126/science.aaz2463} \BibitemShut
  {NoStop}%
\bibitem [{\citenamefont {Del~Pace}\ \emph {et~al.}(2021)\citenamefont
  {Del~Pace}, \citenamefont {Kwon}, \citenamefont {Zaccanti}, \citenamefont
  {Roati},\ and\ \citenamefont {Scazza}}]{delpace2021tunneling}%
  \BibitemOpen
  \bibfield  {author} {\bibinfo {author} {\bibfnamefont {G.}~\bibnamefont
  {Del~Pace}}, \bibinfo {author} {\bibfnamefont {W.~J.}\ \bibnamefont {Kwon}},
  \bibinfo {author} {\bibfnamefont {M.}~\bibnamefont {Zaccanti}}, \bibinfo
  {author} {\bibfnamefont {G.}~\bibnamefont {Roati}}, \ and\ \bibinfo {author}
  {\bibfnamefont {F.}~\bibnamefont {Scazza}},\ }\href {\doibase
  10.1103/PhysRevLett.126.055301} {\bibfield  {journal} {\bibinfo  {journal}
  {Phys. Rev. Lett.}\ }\textbf {\bibinfo {volume} {126}},\ \bibinfo {pages}
  {055301} (\bibinfo {year} {2021})}\BibitemShut {NoStop}%
\bibitem [{\citenamefont {Muniz}\ \emph {et~al.}(2020)\citenamefont {Muniz},
  \citenamefont {Barberena}, \citenamefont {Lewis-Swan}, \citenamefont {Young},
  \citenamefont {Cline}, \citenamefont {Rey},\ and\ \citenamefont
  {Thompson}}]{muniz2020exploring}%
  \BibitemOpen
  \bibfield  {author} {\bibinfo {author} {\bibfnamefont {J.~A.}\ \bibnamefont
  {Muniz}}, \bibinfo {author} {\bibfnamefont {D.}~\bibnamefont {Barberena}},
  \bibinfo {author} {\bibfnamefont {R.~J.}\ \bibnamefont {Lewis-Swan}},
  \bibinfo {author} {\bibfnamefont {D.~J.}\ \bibnamefont {Young}}, \bibinfo
  {author} {\bibfnamefont {J.~R.~K.}\ \bibnamefont {Cline}}, \bibinfo {author}
  {\bibfnamefont {A.~M.}\ \bibnamefont {Rey}}, \ and\ \bibinfo {author}
  {\bibfnamefont {J.~K.}\ \bibnamefont {Thompson}},\ }\href {\doibase
  10.1038/s41586-020-2224-x} {\bibfield  {journal} {\bibinfo  {journal}
  {Nature}\ }\textbf {\bibinfo {volume} {580}},\ \bibinfo {pages} {602}
  (\bibinfo {year} {2020})}\BibitemShut {NoStop}%
\bibitem [{\citenamefont {Lewis-Swan}\ \emph {et~al.}(2021)\citenamefont
  {Lewis-Swan}, \citenamefont {Barberena}, \citenamefont {Cline}, \citenamefont
  {Young}, \citenamefont {Thompson},\ and\ \citenamefont
  {Rey}}]{lewis2021cavity}%
  \BibitemOpen
  \bibfield  {author} {\bibinfo {author} {\bibfnamefont {R.~J.}\ \bibnamefont
  {Lewis-Swan}}, \bibinfo {author} {\bibfnamefont {D.}~\bibnamefont
  {Barberena}}, \bibinfo {author} {\bibfnamefont {J.~R.~K.}\ \bibnamefont
  {Cline}}, \bibinfo {author} {\bibfnamefont {D.~J.}\ \bibnamefont {Young}},
  \bibinfo {author} {\bibfnamefont {J.~K.}\ \bibnamefont {Thompson}}, \ and\
  \bibinfo {author} {\bibfnamefont {A.~M.}\ \bibnamefont {Rey}},\ }\href
  {\doibase 10.1103/PhysRevLett.126.173601} {\bibfield  {journal} {\bibinfo
  {journal} {Phys. Rev. Lett.}\ }\textbf {\bibinfo {volume} {126}},\ \bibinfo
  {pages} {173601} (\bibinfo {year} {2021})}\BibitemShut {NoStop}%
\bibitem [{\citenamefont {Toschi}\ \emph {et~al.}(2005)\citenamefont {Toschi},
  \citenamefont {Barone}, \citenamefont {Capone},\ and\ \citenamefont
  {Castellani}}]{Toschi_2005}%
  \BibitemOpen
  \bibfield  {author} {\bibinfo {author} {\bibfnamefont {A.}~\bibnamefont
  {Toschi}}, \bibinfo {author} {\bibfnamefont {P.}~\bibnamefont {Barone}},
  \bibinfo {author} {\bibfnamefont {M.}~\bibnamefont {Capone}}, \ and\ \bibinfo
  {author} {\bibfnamefont {C.}~\bibnamefont {Castellani}},\ }\href {\doibase
  10.1088/1367-2630/7/1/007} {\bibfield  {journal} {\bibinfo  {journal} {New
  Journal of Physics}\ }\textbf {\bibinfo {volume} {7}},\ \bibinfo {pages} {7}
  (\bibinfo {year} {2005})}\BibitemShut {NoStop}%
\bibitem [{\citenamefont {Bourdel}\ \emph {et~al.}(2004)\citenamefont
  {Bourdel}, \citenamefont {Khaykovich}, \citenamefont {Cubizolles},
  \citenamefont {Zhang}, \citenamefont {Chevy}, \citenamefont {Teichmann},
  \citenamefont {Tarruell}, \citenamefont {Kokkelmans},\ and\ \citenamefont
  {Salomon}}]{bourdel2004experimental}%
  \BibitemOpen
  \bibfield  {author} {\bibinfo {author} {\bibfnamefont {T.}~\bibnamefont
  {Bourdel}}, \bibinfo {author} {\bibfnamefont {L.}~\bibnamefont {Khaykovich}},
  \bibinfo {author} {\bibfnamefont {J.}~\bibnamefont {Cubizolles}}, \bibinfo
  {author} {\bibfnamefont {J.}~\bibnamefont {Zhang}}, \bibinfo {author}
  {\bibfnamefont {F.}~\bibnamefont {Chevy}}, \bibinfo {author} {\bibfnamefont
  {M.}~\bibnamefont {Teichmann}}, \bibinfo {author} {\bibfnamefont
  {L.}~\bibnamefont {Tarruell}}, \bibinfo {author} {\bibfnamefont {S.~J. J.
  M.~F.}\ \bibnamefont {Kokkelmans}}, \ and\ \bibinfo {author} {\bibfnamefont
  {C.}~\bibnamefont {Salomon}},\ }\href {\doibase
  10.1103/PhysRevLett.93.050401} {\bibfield  {journal} {\bibinfo  {journal}
  {Phys. Rev. Lett.}\ }\textbf {\bibinfo {volume} {93}},\ \bibinfo {pages}
  {050401} (\bibinfo {year} {2004})}\BibitemShut {NoStop}%
\bibitem [{\citenamefont {Chen}\ \emph {et~al.}(2005)\citenamefont {Chen},
  \citenamefont {Stajic}, \citenamefont {Tan},\ and\ \citenamefont
  {Levin}}]{CHEN20051}%
  \BibitemOpen
  \bibfield  {author} {\bibinfo {author} {\bibfnamefont {Q.}~\bibnamefont
  {Chen}}, \bibinfo {author} {\bibfnamefont {J.}~\bibnamefont {Stajic}},
  \bibinfo {author} {\bibfnamefont {S.}~\bibnamefont {Tan}}, \ and\ \bibinfo
  {author} {\bibfnamefont {K.}~\bibnamefont {Levin}},\ }\href {\doibase
  https://doi.org/10.1016/j.physrep.2005.02.005} {\bibfield  {journal}
  {\bibinfo  {journal} {Physics Reports}\ }\textbf {\bibinfo {volume} {412}},\
  \bibinfo {pages} {1} (\bibinfo {year} {2005})}\BibitemShut {NoStop}%
\bibitem [{\citenamefont {Biss}\ \emph {et~al.}(2022)\citenamefont {Biss},
  \citenamefont {Sobirey}, \citenamefont {Luick}, \citenamefont {Bohlen},
  \citenamefont {Kinnunen}, \citenamefont {Bruun}, \citenamefont {Lompe},\ and\
  \citenamefont {Moritz}}]{biss2022excitation}%
  \BibitemOpen
  \bibfield  {author} {\bibinfo {author} {\bibfnamefont {H.}~\bibnamefont
  {Biss}}, \bibinfo {author} {\bibfnamefont {L.}~\bibnamefont {Sobirey}},
  \bibinfo {author} {\bibfnamefont {N.}~\bibnamefont {Luick}}, \bibinfo
  {author} {\bibfnamefont {M.}~\bibnamefont {Bohlen}}, \bibinfo {author}
  {\bibfnamefont {J.~J.}\ \bibnamefont {Kinnunen}}, \bibinfo {author}
  {\bibfnamefont {G.~M.}\ \bibnamefont {Bruun}}, \bibinfo {author}
  {\bibfnamefont {T.}~\bibnamefont {Lompe}}, \ and\ \bibinfo {author}
  {\bibfnamefont {H.}~\bibnamefont {Moritz}},\ }\href {\doibase
  10.1103/PhysRevLett.128.100401} {\bibfield  {journal} {\bibinfo  {journal}
  {Phys. Rev. Lett.}\ }\textbf {\bibinfo {volume} {128}},\ \bibinfo {pages}
  {100401} (\bibinfo {year} {2022})}\BibitemShut {NoStop}%
\bibitem [{\citenamefont {Mazza}\ \emph {et~al.}(2021)\citenamefont {Mazza},
  \citenamefont {Amaricci},\ and\ \citenamefont
  {Capone}}]{mazza_interface_BCSBEC_2021}%
  \BibitemOpen
  \bibfield  {author} {\bibinfo {author} {\bibfnamefont {G.}~\bibnamefont
  {Mazza}}, \bibinfo {author} {\bibfnamefont {A.}~\bibnamefont {Amaricci}}, \
  and\ \bibinfo {author} {\bibfnamefont {M.}~\bibnamefont {Capone}},\ }\href
  {\doibase 10.1103/PhysRevB.103.094514} {\bibfield  {journal} {\bibinfo
  {journal} {Phys. Rev. B}\ }\textbf {\bibinfo {volume} {103}},\ \bibinfo
  {pages} {094514} (\bibinfo {year} {2021})}\BibitemShut {NoStop}%
\bibitem [{\citenamefont {Sentef}\ \emph {et~al.}(2017)\citenamefont {Sentef},
  \citenamefont {Tokuno}, \citenamefont {Georges},\ and\ \citenamefont
  {Kollath}}]{sentef2017theory}%
  \BibitemOpen
  \bibfield  {author} {\bibinfo {author} {\bibfnamefont {M.~A.}\ \bibnamefont
  {Sentef}}, \bibinfo {author} {\bibfnamefont {A.}~\bibnamefont {Tokuno}},
  \bibinfo {author} {\bibfnamefont {A.}~\bibnamefont {Georges}}, \ and\
  \bibinfo {author} {\bibfnamefont {C.}~\bibnamefont {Kollath}},\ }\href
  {\doibase 10.1103/PhysRevLett.118.087002} {\bibfield  {journal} {\bibinfo
  {journal} {Phys. Rev. Lett.}\ }\textbf {\bibinfo {volume} {118}},\ \bibinfo
  {pages} {087002} (\bibinfo {year} {2017})}\BibitemShut {NoStop}%
\bibitem [{\citenamefont {Mazza}(2017)}]{mazza2017fromsudden}%
  \BibitemOpen
  \bibfield  {author} {\bibinfo {author} {\bibfnamefont {G.}~\bibnamefont
  {Mazza}},\ }\href {\doibase 10.1103/PhysRevB.96.205110} {\bibfield  {journal}
  {\bibinfo  {journal} {Phys. Rev. B}\ }\textbf {\bibinfo {volume} {96}},\
  \bibinfo {pages} {205110} (\bibinfo {year} {2017})}\BibitemShut {NoStop}%
\bibitem [{\citenamefont {Seibold}\ and\ \citenamefont
  {Lorenzana}(2020)}]{seibold2020nonequilibrium}%
  \BibitemOpen
  \bibfield  {author} {\bibinfo {author} {\bibfnamefont {G.}~\bibnamefont
  {Seibold}}\ and\ \bibinfo {author} {\bibfnamefont {J.}~\bibnamefont
  {Lorenzana}},\ }\href {\doibase 10.1103/PhysRevB.102.144502} {\bibfield
  {journal} {\bibinfo  {journal} {Phys. Rev. B}\ }\textbf {\bibinfo {volume}
  {102}},\ \bibinfo {pages} {144502} (\bibinfo {year} {2020})}\BibitemShut
  {NoStop}%
\bibitem [{\citenamefont {Ojeda~Collado}\ \emph {et~al.}(2019)\citenamefont
  {Ojeda~Collado}, \citenamefont {Usaj}, \citenamefont {Lorenzana},\ and\
  \citenamefont {Balseiro}}]{ojeda2019fate}%
  \BibitemOpen
  \bibfield  {author} {\bibinfo {author} {\bibfnamefont {H.~P.}\ \bibnamefont
  {Ojeda~Collado}}, \bibinfo {author} {\bibfnamefont {G.}~\bibnamefont {Usaj}},
  \bibinfo {author} {\bibfnamefont {J.}~\bibnamefont {Lorenzana}}, \ and\
  \bibinfo {author} {\bibfnamefont {C.~A.}\ \bibnamefont {Balseiro}},\ }\href
  {\doibase 10.1103/PhysRevB.99.174509} {\bibfield  {journal} {\bibinfo
  {journal} {Phys. Rev. B}\ }\textbf {\bibinfo {volume} {99}},\ \bibinfo
  {pages} {174509} (\bibinfo {year} {2019})}\BibitemShut {NoStop}%
\bibitem [{\citenamefont {Mazza}\ and\ \citenamefont
  {Fabrizio}(2012)}]{mazza2012dynamical}%
  \BibitemOpen
  \bibfield  {author} {\bibinfo {author} {\bibfnamefont {G.}~\bibnamefont
  {Mazza}}\ and\ \bibinfo {author} {\bibfnamefont {M.}~\bibnamefont
  {Fabrizio}},\ }\href {\doibase 10.1103/PhysRevB.86.184303} {\bibfield
  {journal} {\bibinfo  {journal} {Phys. Rev. B}\ }\textbf {\bibinfo {volume}
  {86}},\ \bibinfo {pages} {184303} (\bibinfo {year} {2012})}\BibitemShut
  {NoStop}%
\bibitem [{\citenamefont {Seibold}\ \emph {et~al.}(2022)\citenamefont
  {Seibold}, \citenamefont {Castellani},\ and\ \citenamefont
  {Lorenzana}}]{seibold2022adiabatic}%
  \BibitemOpen
  \bibfield  {author} {\bibinfo {author} {\bibfnamefont {G.}~\bibnamefont
  {Seibold}}, \bibinfo {author} {\bibfnamefont {C.}~\bibnamefont {Castellani}},
  \ and\ \bibinfo {author} {\bibfnamefont {J.}~\bibnamefont {Lorenzana}},\
  }\href {\doibase 10.1103/PhysRevB.105.184513} {\bibfield  {journal} {\bibinfo
   {journal} {Phys. Rev. B}\ }\textbf {\bibinfo {volume} {105}},\ \bibinfo
  {pages} {184513} (\bibinfo {year} {2022})}\BibitemShut {NoStop}%
\bibitem [{\citenamefont {Collado}\ \emph {et~al.}(2022)\citenamefont
  {Collado}, \citenamefont {Defenu},\ and\ \citenamefont
  {Lorenzana}}]{collado2205}%
  \BibitemOpen
  \bibfield  {author} {\bibinfo {author} {\bibfnamefont {H.~P.~O.}\
  \bibnamefont {Collado}}, \bibinfo {author} {\bibfnamefont {N.}~\bibnamefont
  {Defenu}}, \ and\ \bibinfo {author} {\bibfnamefont {J.}~\bibnamefont
  {Lorenzana}},\ }\href {\doibase 10.48550/ARXIV.2205.06826} {\enquote
  {\bibinfo {title} {Engineering higgs dynamics by spectral singularities},}\ }
  (\bibinfo {year} {2022})\BibitemShut {NoStop}%
\bibitem [{\citenamefont {Breuer}\ and\ \citenamefont
  {Petruccione}(2007)}]{breuerPetruccione2007}%
  \BibitemOpen
  \bibfield  {author} {\bibinfo {author} {\bibfnamefont {H.~P.}\ \bibnamefont
  {Breuer}}\ and\ \bibinfo {author} {\bibfnamefont {F.}~\bibnamefont
  {Petruccione}},\ }\href {\doibase 10.1093/acprof:oso/9780199213900.001.0001}
  {\emph {\bibinfo {title} {The {{Theory}} of {{Open Quantum Systems}}}}},\
  \bibinfo {edition} {1st}\ ed.,\ Vol.\ \bibinfo {volume} {9780199213}\
  (\bibinfo  {publisher} {{OUP Oxford}},\ \bibinfo {year} {2007})\BibitemShut
  {NoStop}%
\bibitem [{\citenamefont {Weimer}(2015)}]{weimer2015variational}%
  \BibitemOpen
  \bibfield  {author} {\bibinfo {author} {\bibfnamefont {H.}~\bibnamefont
  {Weimer}},\ }\href {\doibase 10.1103/PhysRevLett.114.040402} {\bibfield
  {journal} {\bibinfo  {journal} {Phys. Rev. Lett.}\ }\textbf {\bibinfo
  {volume} {114}},\ \bibinfo {pages} {040402} (\bibinfo {year}
  {2015})}\BibitemShut {NoStop}%
\bibitem [{SM()}]{SM}%
  \BibitemOpen
  \href@noop {} {\enquote {\bibinfo {title} {See supplemental material at [url
  will be inserted by publisher] for (i) the derivation of the bcs dissipative
  dynamics by using a time dependent variational principle for the density
  matrix, (ii) the discussion of the long-time dynamics after a quench of
  dissipation and (iii) the frequency renormalization after a double quench
  through a soliton solution.}}\ }\BibitemShut {NoStop}%
\end{thebibliography}
%

%

\widetext
\clearpage

\setcounter{equation}{0}
\setcounter{figure}{0}
\setcounter{table}{0}
\setcounter{page}{1}
\renewcommand{\theequation}{S\arabic{equation}}
\setcounter{figure}{0}
\renewcommand{\thefigure}{S\arabic{figure}}
\renewcommand{\thepage}{S\arabic{page}}
\renewcommand{\thesection}{S\arabic{section}}
\renewcommand{\thetable}{S\arabic{table}}
\makeatletter

\renewcommand{\thesection}{\arabic{section}}
\renewcommand{\thesubsection}{\thesection.\arabic{subsection}}
\renewcommand{\thesubsubsection}{\thesubsection.\arabic{subsubsection}}

\def\bcen{\begin{center}}
\def\ecen{\end{center}}

\def\a{\alpha}       \def\b{\beta}   \def\g{\gamma}   \def\d{\delta} 
\def\e{\varepsilon}  \def\z{\zeta}   \def\h{\eta}     \def\th{\theta}
\def\k{\kappa}       \def\l{\lambda} \def\m{\mu}      \def\n{\nu}
\def\x{\xi}          \def\p{\pi}     \def\r{\rho}     \def\s{\sigma}
\def\t{\tau}         \def\f{\varphi} \def\ph{\varphi} \def\c{\chi}
\def\ps{\pi}        \def\y{\upsilon}\def\o{\omega}   \def\si{\varsigma}
\def\G{\Gamma}       \def\D{\Delta}  \def\Th{\Theta}  \def\L{\Lambda}  
\def\X{\Xi}          \def\P{\Pi}     \def\Si{\Sigma}  \def\F{\Phi}    
\def\Ps{\Psi}        \def\O{\Omega}  \def\Y{\Upsilon} \def\lg{\langle}

\def\PP{{\cal P}}\def\EE{{\cal E}}\def\MM{{\cal M}} \def\VV{{\cal V}}
\def\CC{{\cal C}}\def\FF{{\cal F}}\def\HH{{\cal H}}\def\WW{{\cal W}}
\def\TT{{\cal T}}\def\NN{{\cal N}}\def\BB{{\cal B}} \def\II{{\cal I}}
\def\RR{{\cal R}}\def\LL{{\cal L}}\def\JJ{{\cal J}} \def\OO{{\cal O}}
\def\DD{{\cal D}}
\def\AAA{{\cal A}}
\def\GG{{\cal G}} \def\SS{{\cal S}}
\def\ZZ{{\cal Z}} \def\UU{{\cal U}}
\def\SB{{\cal S}{\cal B}}
\def\aa{{\V \a}}
\def\hh{{\V h}}\def\HHH{{\V H}}
\def\nn{{\V \n}}\def\pp{{\V p}}\def\mm{{\V m}}\def\qq{{\bf q}}
\def\RRR{\mathbb{R}} \def\CCC{\mathbb{C}} \def\NNN{\mathbb{N}}
\def\ZZZ{\mathbb{Z}} 
\def\QQQ{\hbox{\msytw Q}}

\def\AA{\buildrel_{\circ}\over{\mathrm{A}}}

\def\rg{\rangle}
 \def\ul{\underline}
\def\eg{\mbox{\it e.g.\ }}  \def\ie{\mbox{\it i.e.\ }}
\def\=={\equiv} \def\defi{{\buildrel def \over =}}
\def\lft{\left} \def\rgt{\right} \def\dpr{\partial} \def\der{{\rm d}}
\def\us{\underline \s} \def\ue{{\underline \e}} \def\la{\left\langle}
\def\ra{\right\rangle} 
\def\qed{\raise1pt\hbox{\vrule height5pt width5pt depth0pt}}
\def\iome{i\omega_n} \def\iom{i\omega} \def\iom#1{i\omega_{#1}}
\def\iomn{i\omega_n}
\def\epsk{\epsilon({\bf k})} \def\Ga{\Gamma_{\alpha}}
\def\Seff{S_{eff}}  \def\dinf{$d\rightarrow\infty\,$}
\def\cG0{{\cal G}_0} 
\def\cG{{\cal G}}  \def\cU{{\cal U}}  \def\cS{{\cal S}}
\def\divnum{\frac{1}{N_s}} \def\vac{|\mbox{vac}\rangle}
\def\intR{\int_{-\infty}^{+\infty}} \def\intbeta{\int_{0}^{\beta}}
\def\spinup{\uparrow} \def\spindown{\downarrow} 
\def\up{\uparrow} \def\down{\downarrow} \def\dw{\downarrow}
\def\vk{{\bf k}} \def\qa{{\bf q}} \def\vQ{{\bf Q}}
\def\bk{{\bf k}}\def\bR{{\bf R}}
\def\bq{{\bf q}}
\def\ka{{\bf k} \alpha} 
\def\vr{{\bf r}} \def\q{{\bf q}}  \def\R{{\bf R}}  \def\vR{{\bf R}}
\def\Ak{{\bf A}} \def\Akt{{\bf A}(t)} \def\Ek{{\mathbf E}}
\def\kp{\bbox{k'}} \def\hc{\mbox{h.c.}} \def\Im{\mbox{Im}}
\def\ie{\hbox{\it i.e.\ }} \def\eg{\mbox{\it e.g.\ }}
\def\ie{\mbox{\it i.e.\ }} \def\=={\equiv}
\def\defi{{\buildrel def \over =}} \def\nt{\widetilde{n}}
\def\Im{{\rm Im}} \def\Re{{\rm Re}} \def\Tr{{\rm Tr}\,}
\def\det{{\rm det}\,} \def\ep0{\epsilon_{p}} \def\ed0{\epsilon_{f}}
\def\tpd{V_{fp}} \def\unmezzo{\frac{1}{2}}
\def\ispin{\{\underline{s}\}}
\def\ispinp{\{\underline{s'}\}}

\def\dt{\Delta \tau}

\def\be{\begin{equation}}
\def\ee{\end{equation}}

\def\cc{c^{\dagger}}
\def\ca{c^{\phantom{\dagger}}}
\def\dc{d^{\dagger}}
\def\da{d^{\phantom{\dagger}}}
\def\ac{a^{\dagger}}
\def\aa{a^{\phantom{\dagger}}}

\def\tc{\tilde{c}^{\dagger}}
\def\ta{\tilde{c}^{\phantom{\dagger}}}

\def\abcd{\alpha \beta \gamma \delta}

\renewcommand{\thefigure}{S\arabic{figure}}
\def\pf{\psi}
\def\pfs{\psi^*}

\def\pb{\phi}
\def\pbs{\phi^*}

\def\Hsys{H_{\text{sys}}}
\def\Hbath{H_{\text{bath}}}
\def\Hsb{H_{\text{s-b}}}
\def\rsb{\rho_{I}}
\def\drsb{\dot{\rho}_{I}(t)}


\def\sk{\sigma_{\bk}}

\begin{center}
\large{\bf Supplemental Material to 
`Dissipative Dynamics of a 
Fermionic Superfluid with Two-Body Losses' \\}
\end{center}

\author{Giacomo Mazza}
\affiliation{Dipartimento di Fisica dell'Universit\`a di Pisa, Largo Bruno Pontecorvo 3, I-56127 Pisa, Italy}
\affiliation{Department of Quantum Matter Physics, University of Geneva, Quai Ernest-Ansermet 24, 1211 Geneva, Switzerland}

\author{Marco Schir\`o}
\affiliation{JEIP, UAR 3573 CNRS, Coll\`{e}ge de France, PSL Research University, 11 Place Marcelin Berthelot, 75321 Paris Cedex 05, France}

\maketitle

In this Supplemental Material, we provide details on (i) the derivation
of the BCS dissipative dynamics by using a time dependent variational 
principle for the density matrix, (ii) the long-time behavior of density and order parameter after
a quench of dissipation and (iii) the 
{dissipative soliton solution leading for the double quench dynamics.}

\section{BCS Dissipative Dynamics from Variational Principle}
\label{sec:variational}

The time evolution of a density matrix according to a given Liouvillian 
reads
\begin{equation}
i\dot{\rho} = {\cal L}[\rho]
\label{eq:dot_rho}.
\end{equation}
Eq.~\ref{eq:dot_rho} can be cast in term of a variational principle, by 
introducing an auxiliary density matrix $\rho_{\mathrm{aux}}$, and requiring 
stationarity of the functional 
\begin{equation}
{\cal S}[\rho,\rho_{\mathrm{aux}}] = \int d t \Tr \left[ \rho_\mathrm{aux} \left( i \dot{\rho} - {\cal L}[\rho] \right) \right] 
\qquad 
\frac{\d {\cal S}[\rho,\rho_{\mathrm{aux}}] }{\d \rho_{\mathrm{aux}}}=0.
\label{eq:Sfunctional}
\end{equation}
We now consider a density matrix of the BCS type, $\rho = \rho_0$, and compute the functional~\ref{eq:Sfunctional}  for generic auxiliary 
density matrix $\rho_{\mathrm{aux}}$.
This is straightforwardly computed by using Wick 
theorem. In particular, for any operator $O$, 
a trace of the type $\mathrm{Tr} \left( \rho_0 \rho_{\mathrm{aux}} O \right) $ can be expressed as
\begin{equation}
\mathrm{Tr} \left( \rho_0 \rho_{\mathrm{aux}} O \right) = 
 \sum_{contractions} \Tr \left[ \rho_0 \left(  \underbrace{\rho_{\mathrm{aux}} O} \right) \right],
 \label{eq:wick_example}
\end{equation}
where the symbol $\left(  \underbrace{\rho_{\mathrm{aux}} O} \right)$ means 
the contractions using single fermionic lines of the operator $\rho_{\mathrm{aux}} O.$
By further singling out the contractions only involving terms
terms in the operator $O$, it is possible to reconstruct the expectation value 
\ref{eq:wick_example}
in terms of  the contractions of the operator $O$ times the expectation value 
of the non-contracted part of $\rho_\mathrm{aux}O$,
\begin{equation}
\mathrm{Tr} \left( \rho_0 \rho_{\mathrm{aux}} O \right) = 
\sum_{contractions} \Tr \left( \rho_0 \underbrace{O} \right) \Tr \left( \rho_0  \underbrace{\rho_{\mathrm{aux}} 
\d O} \right),
\label{eq:wick_contractions}
\end{equation}
where $\d O$ indicates the part of $O$ not included in the contraction 
$\Tr \left( \rho_0 \underbrace{O} \right).$
By applying \ref{eq:wick_contractions} to the dissipator in the main 
text, and contracting in both the normal and anomalous channels, we get
\begin{equation}
\begin{split}
{\cal L}[\rho_0,\rho_{\mathrm{aux}}] & 
= 
\Tr \left\lbrace \rho_{\mathrm{aux}} [H_{BCS},\rho_0]\right\rbrace +
\left( 
i \G \D \Tr\left\lbrace  \rho_{\mathrm{aux}} [\ca_{i \downarrow} \ca_{i \uparrow},\rho_0]\right\rbrace
- i \G  \D^* \Tr\left\lbrace  \rho_{\mathrm{aux}} [\cc_{i \uparrow} \cc_{i \downarrow},\rho_0]\right\rbrace \right) \\
& 
- i \G 	\frac{n}{2} 
\sum_{\s}
\Tr\left\lbrace  \rho_{\mathrm{aux}}
{\cal L}_{\s}^{\mathrm{1p-loss}}[\rho_0]
\right \rbrace
\end{split}
\label{eq:trace_lindblad}
\end{equation}
where  $H_{BCS}$ is the unitary BCS Hamiltonian with 
pairing $-|U|$, i.e.
\begin{equation}
H_{BCS} =  \sum_{<ij>\s} t_{ij} \cc_{i \s} \ca_{j\s}- |U| \left(\D \ca_{i \down} \ca_{i \up} +\D^* \cc_{i \up} \cc_{i \down}\right),
\end{equation}
and $\Delta = \mathrm{Tr} \left( \rho_0 \cc_{i \uparrow} \cc_{i \downarrow} \right)$.
Plugging \ref{eq:trace_lindblad} into the variational 
principle \ref{eq:Sfunctional} we get the 
variational dynamics for $\rho_0$ reported in Eq.~(3)
of the main text, i.e. a dynamics for the Anderson pseudospins which read
\begin{align}
\dot{\sigma}_{\bk}^x &=-2\varepsilon_{\bk}\sigma^y_{\bk}+2\mbox{Im}(\Phi)\sigma^z_{\bk}-\Gamma n \sigma^x_{\bk}\\
\dot{\sigma}_{\bk}^y &=2\varepsilon_{\bk}\sigma^x_{\bk}-2\mbox{Re}(\Phi)\sigma^z_{\bk}-\Gamma n \sigma^y_{\bk}\\
\dot{\sigma}_{\bk}^z &=2\mbox{Re}(\Phi)\sigma^y_{\bk}-2\mbox{Im}(\Phi)\sigma^x_{\bk}-\Gamma n\left(\sigma^z_{\bk}+1\right)
\end{align}
with $\Phi=(-\vert U\vert+i\Gamma)\Delta$. We emphasize that the density matrix  $\rho_0$ is a Gaussian BCS mixed state with a definite value of the superfluid order parameter. We note that for our variational approach to work we do not need to write down the explicit form of this state, rather we only need to know that this state satisfies Wick's theorem (with normal and anomalous averages) to obtain the Lindblad master equation in Eq.(3) of the main text.

\section{Dynamics of Particle Density and Pseudo-Spin Length}

Here we derive Eq.(8) of the main text, describing the dynamics of particle density in presence of two-body losses. Using the definition of the density and the variational dynamics of Anderson's pseudo-spins we have
\begin{equation}
\dot{n}=\frac{1}{V}\sum_{\bk}\dot{\sigma^z_{\bk}}
=-\frac{\Gamma n}{V}\sum_{\bk}\left(1+\sigma^z_{\bk}\right)+
\frac{2}{V}\sum_{\textbf{k}}\left(\mbox{Re}\Phi \sigma^y_{\bk}-\mbox{Im}\Phi \sigma^x_{\bk}\right)
\end{equation}
We can rewrite the right hand side in term of the density and the order parameter $\Delta$, using the fact that
\begin{align}
\Delta&=\frac{1}{V}\sum_{\bk} \left(\sigma^{x}_{\bk}+i\sigma^y_{\bk}\right)\\
\Phi&=(-\vert U\vert+i\Gamma)\Delta
\end{align}
to finally obtain 
\begin{equation}\label{eqn:dotn}
\dot{n}=-\Gamma n^2-2\Gamma\vert\Delta\vert^2
\end{equation}
An explicit solution for the particle density can be obtained easily for $\vert\Delta\vert =0$ which gives
\begin{align}
n(t)=\frac{n_0}{1+n_0\Gamma t}
\end{align}
namely a power-law decay of the density, $n~1/t$. Simlarly we can write down an equation of motion for the length of the pseudo-spins as
\begin{equation}
\mathcal{L}=\frac{1}{V}\sum_{\textbf{k}\alpha}\left(\sigma^{\alpha}_{\bk}\right)^2
\end{equation}
By taking the time-derivative and using the equations of motion in the main text we obtain
\begin{equation}
\dot{\mathcal{L}}=\frac{2}{V}\sum_{\textbf{k}\alpha}\left(\sigma^{\alpha}_{\bk}\right)\dot{\sigma^{\alpha}_{\bk}}=-2\Gamma n \left(\mathcal{L}+
(n-1)\right)
\label{eq:eom_pseudospin_norm}
\end{equation}
where we have used  the definition of the time-dependent density $n=1+\frac{1}{V}\sum_{\bk}\sigma^z_{\bk}$.  We see that the length of the pseudospin is not conserved in presence of dissipation, $\Gamma\neq0$. Here we have used the fact that the coherent part of the evolution preserves the pseudo-spin length, even in presence of an imaginary coupling in the pairing field $\Phi(t)=(-\vert U\vert+i\Gamma)\Delta$, and that only the effective single particle losses $\Gamma_{eff}=\Gamma n$ included in our variational treatment give rise to a pseudo-spin length decay as opposed to the description of Ref~\cite{yamamoto2021collective}. We note that the decay of the length $\mathcal{L}$ is only controlled by the dynamics of the density.
For a particle density decaying as $n(t)=n_0/(1+n_0\Gamma t) $ we obtain
\begin{align}
\mathcal{L}(t)=\frac{2\Gamma n_0}{(1+n_0\Gamma t)^2}\left((1-n_0)t+\Gamma n_0 t^2/2+\frac{1}{2\Gamma n_0}\right)
\end{align}
For half-filling initial state we have $\mathcal{L}(t)=1-2\Gamma t/(1+\Gamma t)^2$, namely the pseudo-spin length rapidly deviates from one and then recovers it at long times in a power law fashion $\mathcal{L}(t)\sim 1-2/\Gamma t$.

\section{Comparison with Single Particle Losses and Role of Additional Single-Particle Pump}
 
{In this section} we provide further details on the {dissipative} dynamics 
in {the case of (i) dissipation due to single particle losses and (ii) the case where 
two-body losses are counterbalanced with a single particle pump term.} 

In the case of losses purely due to single-particle terms we can write the Lindblad master equation for the system as in Eq. (1) of the main text,
with the same Hamiltonian part and with local jump operators given by $L_i =\sqrt{\G_{1}} \ca_{i \s} $, where $\G_{1}$ is the rate of single-particle loss.
Applying the above variational treatment  we obtain the following equations of motion for the Anderson pseudo-spins
\begin{align}\label{eq:loss1_sigmax}
\dot{\sigma}_{\bk}^x &=-2\varepsilon_{\bk}\sigma^y_{\bk}-2\vert U\vert \mbox{Im}(\Delta)\sigma^z_{\bk}-\Gamma_1 \sigma^x_{\bk}\\
\label{eq:loss1_sigmay}
\dot{\sigma}_{\bk}^y &=\phantom{-}2\varepsilon_{\bk}\sigma^x_{\bk}+2\vert U\vert\mbox{Re}(\Delta)\sigma^z_{\bk}-\Gamma_1 \sigma^y_{\bk}\\
\dot{\sigma}_{\bk}^z &=-2\vert U\vert\mbox{Re}(\Delta)\sigma^y_{\bk}+2\vert U\vert\mbox{Im}(\Delta)\sigma^x_{\bk}-\Gamma_1 \left(\sigma^z_{\bk}+1\right)\label{eq:loss2_sigmaz}
\end{align}
We see that, at variance with the case of 
two-body losses discussed before, the rate of dissipation 
$\G_1$   is now independent from the density and thus constant in time. {Moreover, the complex interaction term is absent.}
 {The} closed  equation of motion for the particle density  now reads
\begin{align}
\dot{n}=\frac{1}{V}\sum_{\bk}\dot{\sigma^z_{\bk}}
=-\frac{\Gamma_1}{V}\sum_{\bk}\left(1+\sigma^z_{\bk}\right)-
\frac{2\vert U\vert}{V}\sum_{\textbf{k}}\left(\mbox{Re}\Delta \sigma^y_{\bk}-\mbox{Im}\Delta \sigma^x_{\bk}\right)=-\Gamma_1 n,
\end{align}
namely $n(t)=n_0\exp(-\Gamma t)$.

\begin{figure}
\includegraphics[scale=0.5]{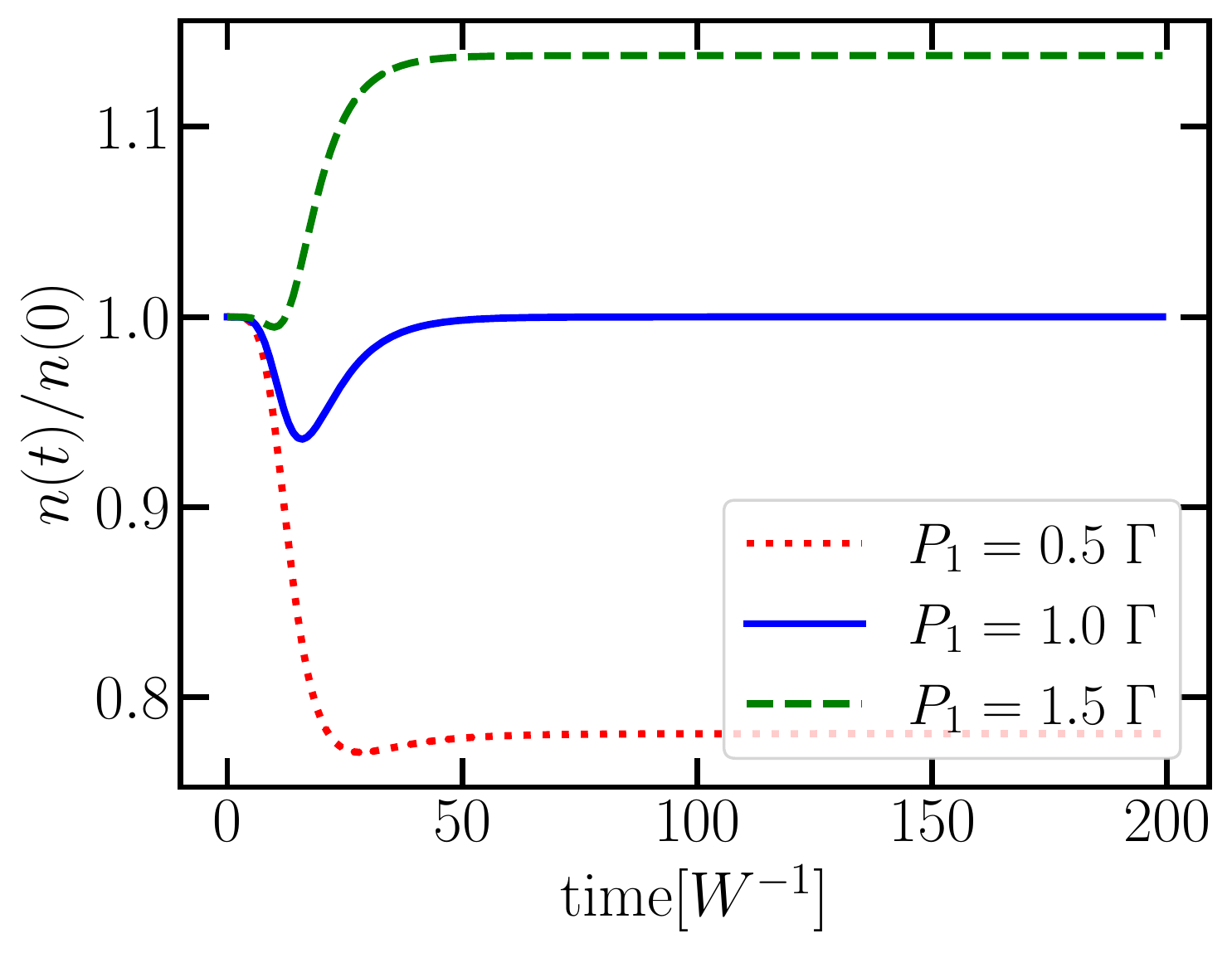}
\caption{Dynamics of density in the case of two-body losses 
balanced by the single-particle pumps for different ratios $\G/P_1$. }
\label{sfig1}
\end{figure}

In the case of single particle pump, the jump operator reads
$L_i =\sqrt{P_{1}} c^{\dagger}_{i \s} $, where $P_{1}$ is the rate of single-particle pump. 
The equations of motion read
 \begin{align}\label{eq:pump1_sigmax}
\dot{\sigma}_{\bk}^x &=
\left(\dot{\sigma}_{\bk}^x\right)_{\rm 2p-loss} -P_1  \sigma^x_{\bk}\\
\label{eq:pump1_sigmay}
\dot{\sigma}_{\bk}^y &=
\left(\dot{\sigma}_{\bk}^y\right)_{\rm 2p-loss} - P_1 \sigma^y_{\bk}\\
\dot{\sigma}_{\bk}^z &=
\left(\dot{\sigma}_{\bk}^z\right)_{\rm 2p-loss}
-P_1 \left(\sigma^z_{\bk}-1\right)\label{eq:pump1_sigmaz}
\end{align}
{where $\left(\dot{\sigma}_{\bk}^{x/y/z}\right)_{\rm 2p-loss}$ are the equations 
of motion reported in the main text.}
{We see that the effect of single particle pump is to introduce a term which re-populates the $\sigma^z_{\bk}$ component and prevents the complete depletion of the system.}
Writing down the dynamics of the particle density we obtain
\begin{equation}
{\dot{n}=-\Gamma n^2-2\Gamma\vert\Delta\vert^2 - P_1 (n-2)}
\end{equation}
{where we see that a steady state solution with $n_{\infty} = n_0 = 1$ 
and $|\D_{\infty}|=0$ is established for $\G=P_1$. For different ratios 
$\G/P_1$ we obtain steady states with $n_{\infty} \neq n_0$ (Fig.~\ref{sfig1}).}

{Based on the above equation of motion, we understand 
the difference between the two-body losses and the the cases of single-particle loss 
or two-body losses supplemented by single-particle pump.}
In both cases each Anderson Pseudo-spin at long-time has an exponential decay in time towards 
the vacuum with a finite rate which does not depend on time,
i.e. $\sigma^{x,y}_{\bk}(t)\sim \exp(-\Gamma_1 t)${, for the single-particle losses,
and $\sigma_{\bk}^{x,y} (t) \sim \exp[-(\Gamma n_{\infty} +P_1) t]$, for the additional single particle pump,} 
as one can obtain from 
the pseudo-spin dynamics disregarding the unitary evolution.  

On top of this exponential decay the unitary part of the evolution gives rise to a momentum dependent phase for each pseudo-spin, which leads to dephasing and decay for the order parameter. To estimate this decay we note that when the order parameter is small we can disregard it from the evolution and obtain an oscillation frequency set by $\varepsilon_{\bk}$ 
which{, after summing over momenta,}
gives rise {to a $1/t$ decay,} 
{\ie $\sum_{\bk} e^{i \varepsilon_{\bk} t} \sim \rho_0/t$ with $\rho_0$ the density of states at the Fermi level.} 
For single-particle losses, the power-law dephasing is however subleading and the overall decay of the order parameter 
is exponential, as shown in Figure 1 of the main text. 

In the case of two-body losses, our variational dynamics gives an evolution of each pseudospin with an effective decay rate $\Gamma_{\rm eff}(t)=\Gamma n(t)$. 
In the long time limit we have $n(t)\sim 1/t$, namely each pseudospin decays as 
$$
\sigma^{x,y}_{\bk}(t)\sim \frac{1}{1+\Gamma n_0 t}\sim 1/t
$$
rather than the exponential decay of single particle losses. 
In addition, including the dephasing of each pseudo-spin from the unitary evolution discussed above we obtain an overall $1/t^2$ decay of the superfluid order parameter, in agreement with our numerical analysis. 
This explains why the order parameter decays {more} slowly than the density.  
Also, this explain that the crossover towards this power law only occurs 
when the density starts to deviate significantly from a constant value, 
while for short times the order parameter decay is exponential.

\section{Frequency Renormalization after a double Quench: Dissipative Solitons}
In this section we briefly review the soliton solution of BCS dynamics~\cite{BarankovLevitovPRL06} and discuss 
its generalization 
to the dissipative case to interpret the results of the double quench of interaction and dissipation.  

In the unitary case the 
 dynamics of the BCS problem reads
\begin{align}\label{eq:sigmax}
\dot{\sigma}_{\bk}^x &=-2\varepsilon_{\bk}\sigma^y_{\bk}\\
\label{eq:sigmay}
\dot{\sigma}_{\bk}^y &=\phantom{-}2\varepsilon_{\bk}\sigma^x_{\bk}+2\vert 
U{_f}\vert
\Delta \sigma^z_{\bk}\\
\dot{\sigma}_{\bk}^z &=-2\vert U{_f} \vert \Delta 
\sigma^y_{\bk}\label{eq:sigmaz},
\end{align}
{where for the unitary case 
the dynamics of the phase of the order parameter 
is frozen and we choose 
$\mbox{Im}\Delta =0 $ and $\mbox{Re}\Delta = \D = \frac{1}{2} \quave{\sk^x}$).} 
We parametrize the pseudo-spin components as
\begin{align}
\sk^x = A_{\bk} \D \qquad
\sk^y =  B_{\bk} \dot{\D} \qquad
\sk^z = C_{\bk} \D^2 - D_{\bk} 
\label{eq:ansatz_soliton}
\end{align}
where the coefficients $A_{\bk},B_{\bk},C_{\bk},D_{\bk} $ have to satisfy the consistency conditions
\begin{align}
\frac{1}{2} \sum_{\bk } A_{\bk} = 1 \qquad
\sum_{\bk} B_{\bk} = 0 \qquad
 \sum_{\bk} C_{\bk} \D^2 - D_{\bk} = 0
 \label{eq:concistency_ansatz}
\end{align}
By plugging 
the ansatz~\ref{eq:ansatz_soliton} 
in the equations of motion, we determine 
$A_{\bk}$ and $B_{\bk}$ as a function of $C_{\bk}$ 
and $D_{\bk}$. 
We therefore impose the conservation of the pseudo-spin
\begin{align}
(\sk^{1})^2 + (\sk^{2})^2 + (\sk^{3})^2 = 1 
\label{eq:pseudospin_conservation}
\end{align}
to obtain a dynamical equation for $\D$ which reads
\begin{equation}
(\dot{\D})^2 + U^2 (\D^2 - \D_{+}^2) (\D^2 - \D_{-}^2) =0
\label{eq:soliton_train_eom}
\end{equation}
where the coefficients $\Delta_{\pm}$ satisfy the conditions
\begin{align}
&\D_{+}^2 + \D_{-}^2 = - \left(  4 \frac{\epsilon_{\bk}^2}{U^2} - 2 \frac{D_{\bk}}{C_{\bk}} \right) 
\qquad \D_{+}^2 \D_{-}^2 = \frac{D_{\bk}^2-1}{C_{\bk}^2}
\label{eq:delta_sumprod}
\end{align}
{Eq.~\ref{eq:soliton_train_eom} defines a soliton train solution of the form}
\begin{equation}\label{eq:dn}
{ \D(\t) = \D_+ {\rm dn} \left( \frac{t}{|U_f|},1-\frac{\D_-^2}{\D_+^2} \right) }
\end{equation}
{with period}
\begin{align*}
{{\cal T} = 2 \frac{K(1-\a^2)}{|U_f| \D_+}}
\qquad
{\a\equiv\frac{\D_-^2}{\D_+^2}}
\end{align*}
{The dimensionless parameters $\D_\pm$ are determined by the two conditions}
\begin{align}
 &
  |U_f| \sum_{\bk} \frac{2 \epsilon_{\bk} \rm sgn \epsilon_{\bk}}
 {\sqrt{ \left[ U_f^2 \left( \D_+^2 + \D_-^2 \right) + 4\epsilon_{\bk}^2 \right]^2 - 4 U_f^4 \D_+^2 \D_-^2}} 
  = 1
   \label{eq:selfsoliton1} \\
&
E_f = \sum_{\bk } 
 \frac{2 U_f^2 \epsilon_{\bk} \rm sgn \epsilon_{\bk}}{\sqrt{ \left[ U_f^2(\D_+^2 + \D_-^2) + 4 \epsilon_{\bk}^2 \right]^2 - 4 U_f^4 \D_+^2 \D_-^2  }}
\left( \frac{\D_+^2-\D_-^2}{2} - 2 \frac{\epsilon_{\bk}^2}{U_f^2} \right)
- |U_f| \D_+^2 
\label{eq:selfsoliton2}
\end{align}
coming, respectively, from the Eqs.~\ref{eq:concistency_ansatz} 
and the conservation of the total energy after the quantum quench $E_f$.

\subsection{Soliton dynamics with small dissipation}
\begin{figure}
\includegraphics[scale=0.5]{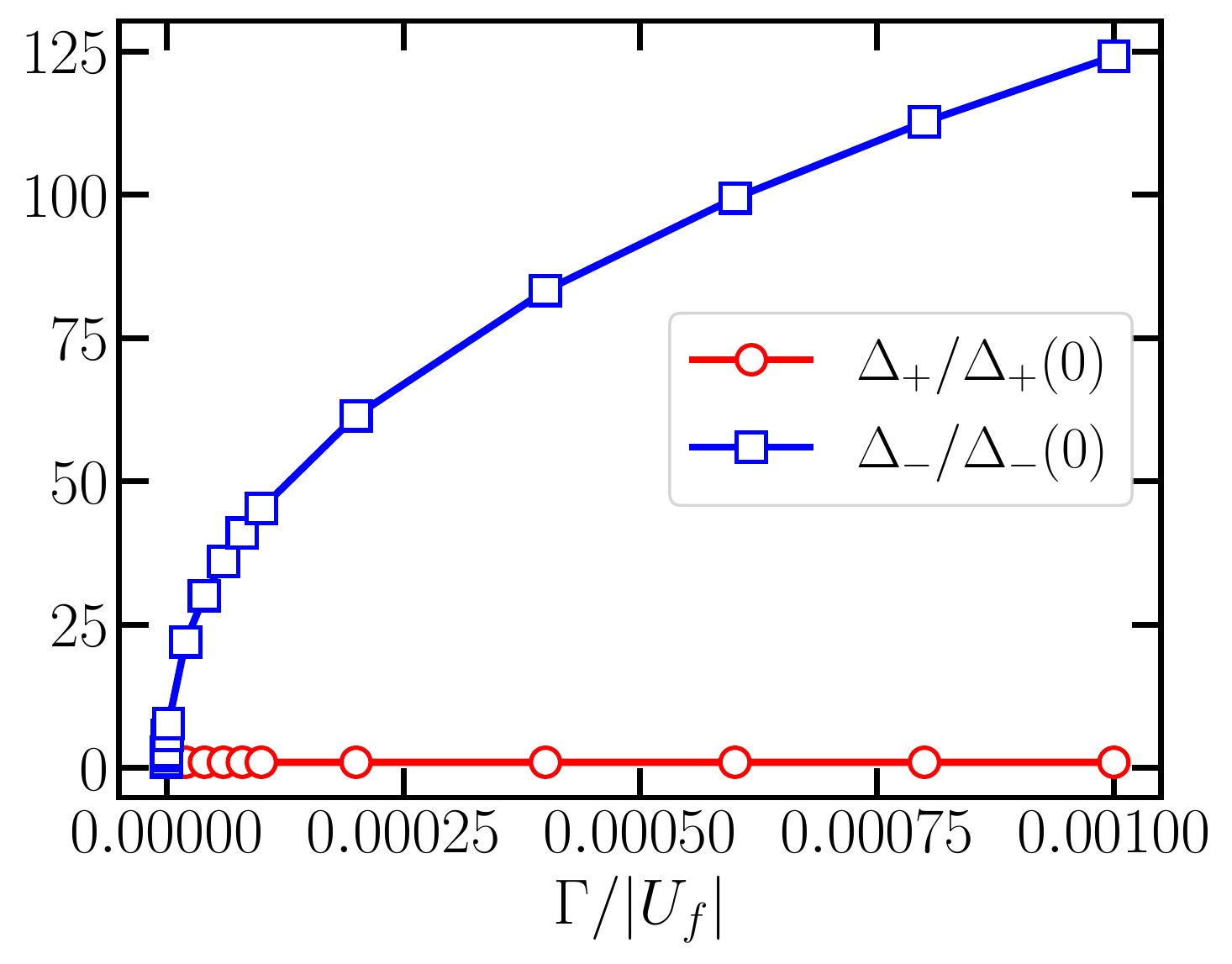}
\includegraphics[scale=0.5]{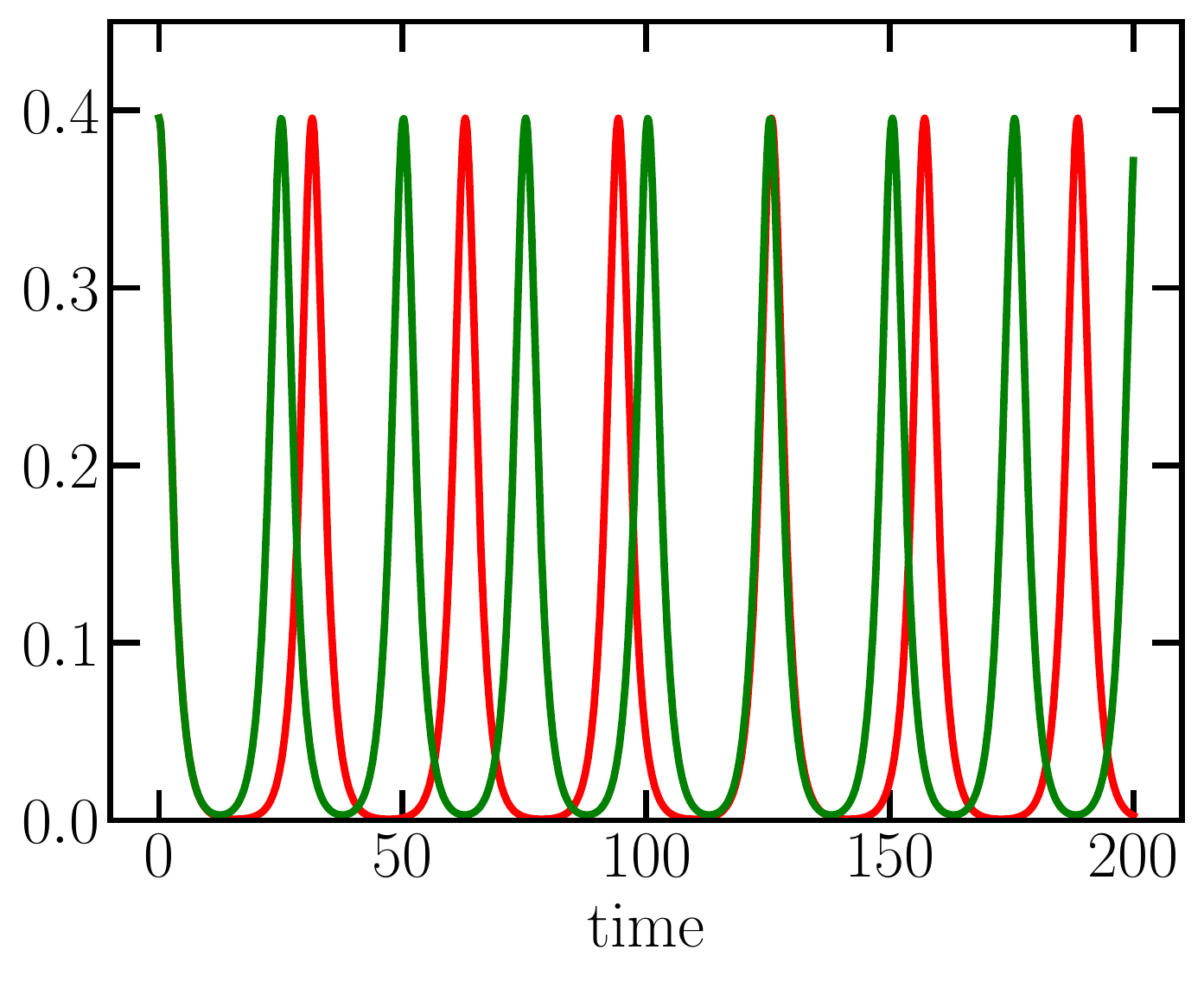}
\caption{(Left) $\D_{\pm}$ parameters as a function of dissipation. 
(Right) dissipative soliton solution, see Eq.~\ref{eq:dn}, for 
$\Gamma/|U_f|=10^{-8}$ (red) and $\G/|U_f|=10^{-6}$ (green).}
\label{sfig2}
\end{figure}

We now extend this treatment to describe the dissipative case. 
We do that by assuming that for $\G \to 0$, 
the main effect of the dissipation is to weakly break 
the conservation of the pseudospin norm. We therefore assume to 
completely drop the $\G$  dependence from the equation of motions 
and include all the effect of the dissipation in a change of the 
pseudospin length 
\begin{equation}
{\cal L}_{\bk} = \sum_{i} \left( \s_{\bk}^i \right)^2 \approx 1 - \d {\cal L}_{\bk}
\end{equation}
At times $\G t \ll 1$ we have from Eq.~\ref{eq:eom_pseudospin_norm}
\begin{align}
{\cal L}_{\bk} \approx 1 - 2 \G \left( t - t_0 \right)
\label{eq:norm_diss}
\end{align}
where $t_0$ is the time at which the weak dissipation 
is switched on {and at short times we have considered $n = 1.$}
In order to get an effective unitary soliton dynamics 
we therefore assume that, on a time scale $t \sim {\cal T}$
corresponding to the first soliton oscillation, 
the pseudospin length can be considered constant and 
estimated using its time averaged value
\begin{align}
{\cal L}_{\bk} \approx
\langle{\cal L}_{\bk} \rangle \equiv 1 - \G {\cal T} 
\label{eq:norm_ansatz}
\end{align}
{where} ${\cal T} $ is the period of the 
soliton train to be determined.
We therefore proceed by parametrizing the pseudo-spins dynamics as 
in the unitary case, and imposing the pseudo-spin conservation Eq.~\ref{eq:norm_diss}~{with a 
reduced pseudo-spin length} 
$$
{(\sk^{1})^2 + (\sk^{2})^2 + (\sk^{3})^2 = 1 - \G {\cal T}}.
$$
{We obtain a dynamics equation equivalent to the one reported 
above with modified equations for the soliton parameters $\D_{\pm}(\G)$}
\begin{align}
 &
  |U_f| \sum_{\bk} \frac{2 \epsilon_{\bk} {\rm sgn} \epsilon_{\bk} }
 {\sqrt{ \left[ U_f^2 \left( \D_+^2 + \D_-^2 \right) + 4\epsilon_{\bk}^2 \right]^2 - 4 U_f^4 \D_+^2 \D_-^2}} 
  = \frac{1}{\sqrt{1 - \frac{2 \G K(1-\a^2)}
  {|U_f| \D_+}}}
   \label{eq:selfsoliton1_diss} \\
&
 \sum_{\bk } 
 \frac{2 U_f^2 \epsilon_{\bk} \rm sgn \epsilon_{\bk} }{\sqrt{ \left[ U_f^2(\D_+^2 + \D_-^2) + 4 \epsilon_{\bk}^2 \right]^2 - 4 U_f^4 \D_+^2 \D_-^2  }}
\left( \frac{\D_+^2-\D_-^2}{2} - 2 \frac{\epsilon_{\bk}^2}{U_f^2} \right)
=
\frac{E_0 + |U_f| \D_+^2 }{\sqrt{1 - \frac{2\G K(1-\a^2)}{|U_f| \D_+}}}
\label{eq:selfsoliton2_diss}
\end{align}
{where we have replaced the period in Eq.~\ref{eq:norm_ansatz} with 
${\cal T} = 2 K(1-\a^2)/U_f\D_+.$}
In the additional Fig.~\ref{sfig2} (left panel)
we show the parameters $\D_{\pm}(\G)$ as a function
of $\G.$ 
We find that $\D_{+}$ remains essentially constant whereas $\D_{-}$ 
increases non-linearly with $\G$. Computing the period of the soliton oscillations we find the logarithmic dependence
that we show in the main text, Figure 3. The key for this effect is therefore the strong non-linear dependence of 
$\D_{-}$ with $\G$ and the fact that the period of oscillation of the soliton train is given by the elliptic integral $\mathcal{T} \sim K(k)$, with $k=1-\D_-^2/\D_+^2$. 
{Increasing $\D_-$ shifts the period away from}
the logarithmic singularity {of the elliptic integral} 
for $k \rightarrow 1$.

\begin{figure}
\includegraphics[scale=0.3]{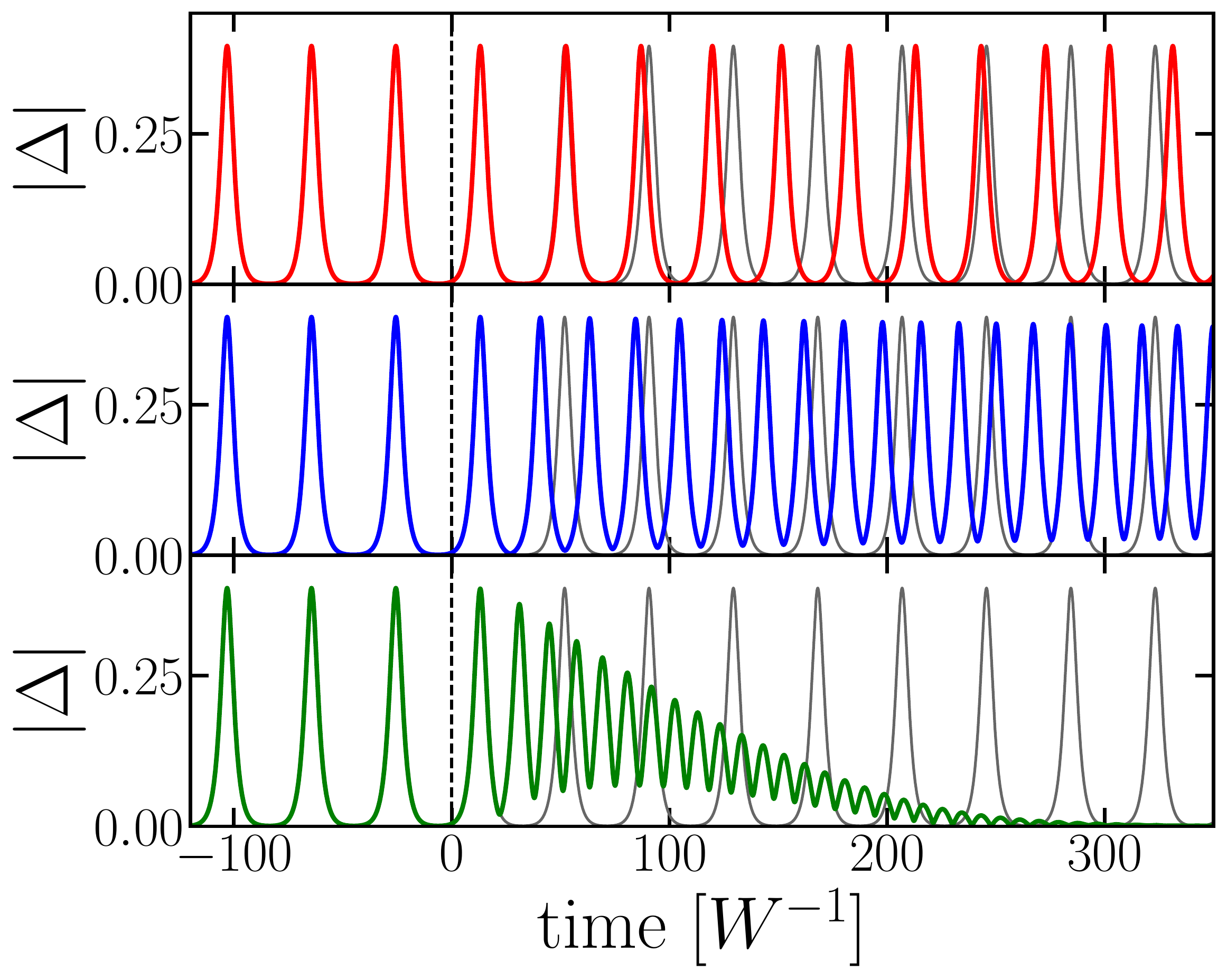}
\caption{Double quench dynamics for in the presence of single-particle losses
for the same parameters of Fig. 2 in the main text.}
\label{sfig3}
\end{figure}

{Eventually, we emphasise that for the dissipative 
soliton solution we considered the density as constant. 
This suggests that the argument for the case of the two-body losses 
can be readily extended to the case of the single-particle losses, 
by considering $\G_1 = \G n \simeq \G.$
As a check we show in Fig.~\ref{sfig3} that the numerical 
solution of the equation of motions in the presence of 
single-particle losses reproduces a phenomenology similar 
to the one presented in the main text for the two-body losses.}

\end{document}